\documentclass[a4paper]{article}

\usepackage[utf8]{inputenc}   
\usepackage[T1]{fontenc}      
\usepackage{lmodern}          

\usepackage[margin=25mm]{geometry}
\usepackage{amsmath, amssymb, amsfonts}
\usepackage{booktabs,array,makecell,longtable,multirow}
\usepackage{graphicx,adjustbox}
\usepackage{caption}
\usepackage{microtype}
\usepackage{verbatim}

\providecommand{\keywords}[1]{%
  \small\textbf{\textit{Keywords---}} #1%
}

\title{Influential scientists shape knowledge flows\\ between science and IGO policy}
\author{
    Kimitaka Asatani$^{1}$,
    Yurie Iwata$^{1}$,
    Yuta Tomokiyo$^{1}$,
    Basil Mahfouz$^{2}$,\\
    Masaru Yarime$^{3}$,
    Ichiro Sakata$^{1}$ \\
    \small $^{1}$The University of Tokyo \\
    \small $^{2}$University College London \\
    \small $^{3}$The Hong Kong University of Science and Technology \\
}
\date{} 

\begin{document}
\maketitle

\begin{abstract}
Intergovernmental organizations (IGOs) increasingly rely on scientific evidence, yet the pathways through which scientific research enters policy remain opaque. By linking 230,737 scientific papers cited in IGO policy documents (2015-2023) to their authors and collaboration networks, we identify a small group of policy-influential scientists (PI-Sci) who dominate this knowledge flow. These scientists form tightly interconnected, internationally spanning co-authorship networks and achieve policy citations shortly after publication, a distinctive feature of cumulative advantage at the science-policy interface. The concentration of influence varies by field: tightly clustered in established domains like climate modeling, and more dispersed in emerging areas like AI governance. Many PI-Sci serve on high-level advisory bodies (e.g., IPCC), and major IGOs frequently co-cite the same PI-Sci papers, indicating synchronized knowledge diffusion through shared expert networks. These findings reveal how network structure and elite brokerage shape the translation of research into global policy, highlighting opportunities to broaden the scope of knowledge that informs policy.
\end{abstract} \hspace{10pt}


\section{Introduction}
Evidence-based policymaking requires effective and bidirectional knowledge exchange between scientists, policymakers, and stakeholders\cite{cash2003knowledge, contandriopoulos2010knowledge, gerber2023bridging}. Intergovernmental organizations (IGOs) such as the Intergovernmental Panel on Climate Change (IPCC) and the World Health Organization (WHO) translate research into guidelines that shape national and multilateral action\cite{haas1992introduction}. However, the mechanisms of science-to-policy translation vary dramatically across domains. Pandemic response demands near-real-time integration of rapidly evolving biomedical findings\cite{yin2021coevolution}, while climate governance relies on the IPCC's periodic assessments to build long-term consensus\cite{gupta2010history}. AI regulation remains nascent, with the UN AI Advisory Body only recently beginning to formalize expert input. Across these contexts, boundary-spanning "interface scientists"\cite{gluckman2021brokerage} broker evidence and influence the direction of policy, prompting discussion about whether scientists should maintain neutrality or embrace advocacy roles\cite{dwivedi2024real, akerlof2022global}.


Influence concentration among elite individuals and institutions is well-researched across creative and scientific arenas \cite{ahmadpoor2017dual,fraiberger2018quantifying}. Large-scale syntheses of Altmetric and Overton records now trace the same stratification at the science-policy nexus \cite{szomszor2022overton}: policy documents disproportionately cite already highly cited articles \cite{bornmann2022relevant,dorta2025kind}, and media coverage further magnifies this skew \cite{dorta2024societal,anderson2020case}. Service on international guideline panels likewise yields reciprocal gains, boosting scientists’ subsequent academic visibility while opening new conduits for research to travel into policy prose \cite{tomokiyo2025researchers,furnas2025partisan}. Yet we still lack a systematic account of the micro-mechanisms that allow individual scholars to convert scholarly capital into policy traction or of whether a narrow “policy elite” dominates attention \cite{van2007rationale,cash2003knowledge}. Clarifying how these influence patterns unfold within the distinct epistemic cultures and governance architectures of domains such as biomedicine, climate science, and artificial intelligence is, therefore, crucial to ongoing debates about equity and representation in evidence-informed decision-making \cite{oliver2014new,parkhurst2017politics}.



Here we analyze 230,737 scientific papers cited in IGO policy documents from 2015 to 2023. By linking Overton records with Scopus author identifiers, we find that authors' prior policy citations are a crucial determinant of future policy uptake. Combined with journal selection and the policy relevance of cited references, these factors largely explain regional disparities in policy citation rates. Then, we identify a small cohort of Policy-Influential Scientists (PI-Sci) who receive a disproportionate share of policy citations, receive them earlier, and collaborate within dense international networks. This concentration is pronounced in mature fields like climate modeling but weaker in rapidly evolving areas like data science. We also document hierarchical and synchronized citation patterns among IGOs, suggesting that limited sets of scientists and organizations disproportionately shape global policy integration. These findings illuminate science-policy dynamics and identify opportunities to enhance inclusive evidence-based decision-making.

\begin{figure*}[!t]
    \centering \includegraphics[width=\linewidth]{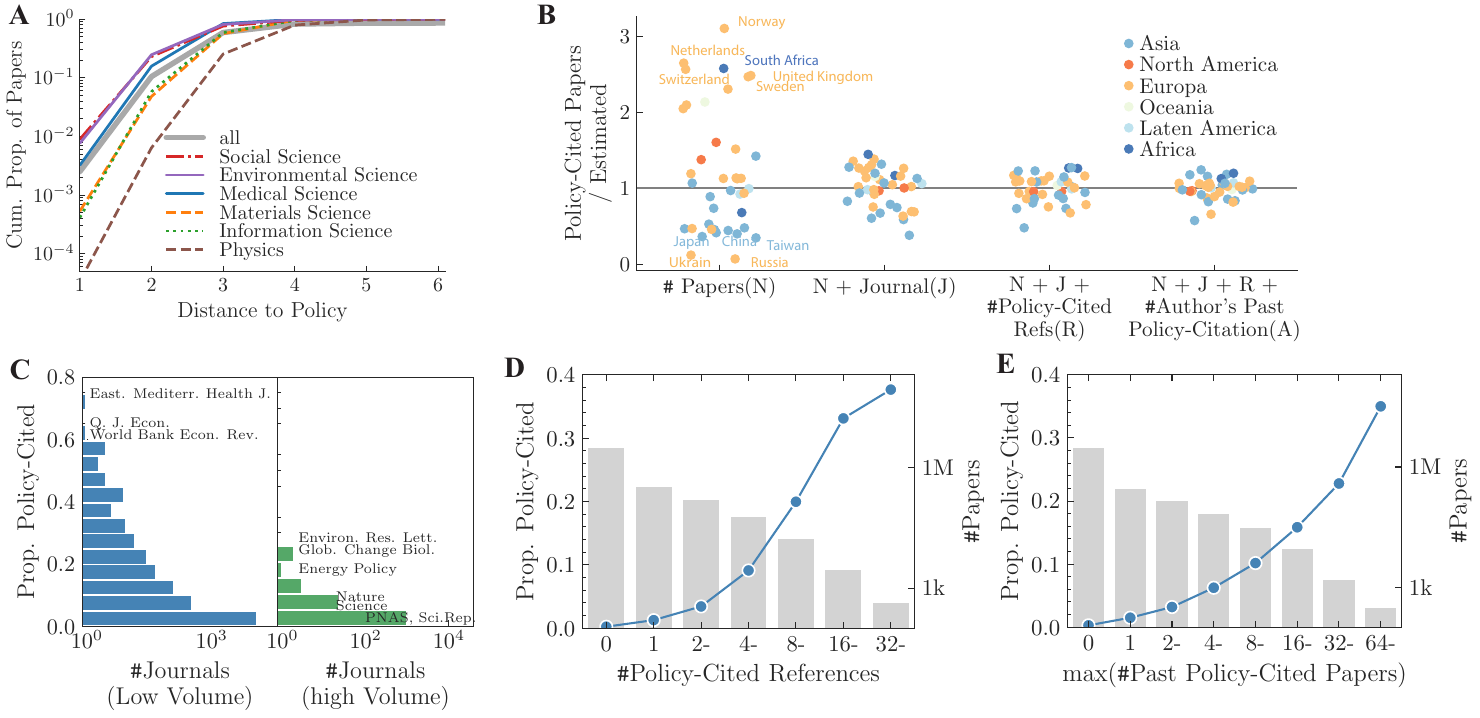}
 \caption{\textbf{Academic Pathways to Policy: Citation Distance and Impact Predictors.}
\textbf{(A)} Cumulative distribution of papers (2015-) by shortest citation-path distance to IGO policy documents.
\textbf{(B)} Ratio of observed to predicted policy-cited papers for 40 most productive countries as predictors are added sequentially: baseline count (N), plus journal (J) shown as N+J, plus number of policy-cited references (R) shown as N+J+R, and plus authors' prior policy-cited papers (A) shown as N+J+R+A.
\textbf{(C)} Policy-citation rates by journal volume: low-volume outlets ($\leq$500 papers in 2020, left) versus high-volume outlets (right). 
\textbf{(D)} Policy-citation probability (blue line, left axis) and paper count (gray bars, right axis) by number of policy-cited references. 
\textbf{(E)} Same as (D) but for a maximum number of authors' prior policy-cited papers.}
    \label{fig01}
    \vspace{-0.3cm}
\end{figure*}

\section{Results}
\subsection{Factors Driving the Policy Impact of Scientific Papers}
Only 0.71\% of papers published since 2015 have been cited directly in IGO documents, yet 93.3\% are connected indirectly through their citation networks (Figure~1A). Across domains, a large proportion of papers are connected to policy within 3 or 4 steps in the citation network, indicating that most papers have some connection to policy. Within three steps, 37.0\% of physics papers and 71.2\% of information science papers connect to at least one IGO citation. These patterns confirm that analyzing directly policy-cited papers (distance = 1) is essential for understanding the broader science-policy ecosystem.

The probability of direct policy citation varies markedly by country. Western and Northern Europe, along with South Africa, display the highest rates, while Eastern Europe and much of East Asia fall below the global average (Figure~1B, first column; full results in Table S1). We also find that the top affiliations by number of policy-cited papers are biased toward European and North American countries (Table S2). To investigate the sources of this disparity, we examined three paper-level attributes (see Methods): journal of publication (J), number of policy-cited references (R), and maximum number of policy-cited papers previously authored by any co-author (A).

Proximity to policy (through journal selection and citation of policy-relevant research) combined with potential academic impact significantly increases the likelihood of future policy citations. Journal publishing patterns partially explain country-level variation in policy citation rates (Figure~1B, second column). While some journals with exceptionally high policy-citation rates (e.g., Quarterly Journal of Economics, 67.5\%) publish fewer than 500 papers annually, high-volume journals—including Environmental Research Letters (24\%), Nature (6.9\%), Science (4.2\%), and PNAS (4.5\%)—substantially exceed the 0.71\% average across all journals (Figure~1C). Policy-cited papers also achieve higher academic citations and greater scholarly impact (Figure S1), corroborating prior findings\cite{yin2021coevolution}. Furthermore, citing policy-relevant research strongly predicts future policy citation: papers citing no policy-cited references have only a 0.25\% chance of subsequent policy citation, while those citing more than eight such papers reach 30\% (Figure~1D). Together, policy proximity and academic impact explain substantial variation in policy citation probability across countries (Figure~1B, third column).

An author's prior policy influence is the final, decisive factor. Policy-citation probability increases sharply with authors' previous policy citations, exceeding 10\% when any co-author has eight or more policy-cited papers (Figure~1E). Incorporating this historical term substantially reduces country-level disparities: 26 of the 40 most productive nations now fall within ±10\% of parity (Figure~1B, rightmost column).  Authors' prior policy influence adjusts for East Asian countries (such as China, Japan, and South Korea) underrepresentation (Table S1), indicating that these countries lack frequently policy-cited scientists. This pattern reveals a pronounced Matthew effect at the science-policy interface, whereby scientists who have previously influenced policy continue to receive disproportionate attention, mirroring cumulative advantage dynamics in academic citation networks\cite{merton1968matthew}.

\begin{figure*}[!t]
    \centering \includegraphics[width=\linewidth]{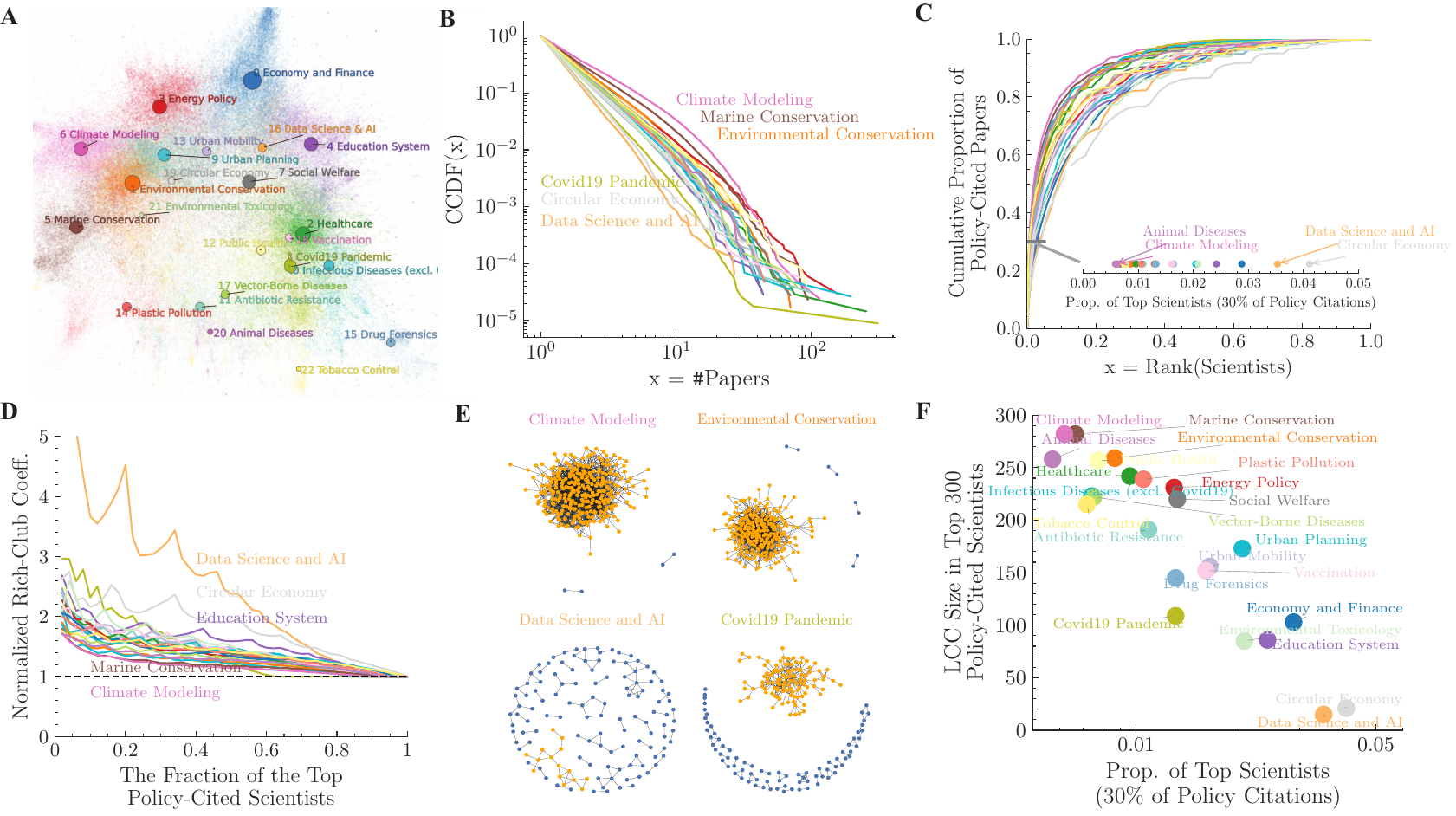}
  \caption{\textbf{Concentration and collaboration among policy-influential scientists.}
\textbf{(A)} Citation network visualization of policy-cited papers colored by 23 data-driven domains. Labels derived from TF-IDF terms in each cluster. Layout calculated using ForceAtlas2\cite{jacomy2014forceatlas2}.
\textbf{(B)} Log-log complementary cumulative distribution of policy-cited papers per scientist for representative domains. 
\textbf{(C)} Cumulative share of policy-cited output plotted against author rank. The dashed line indicates a 30\% threshold defining policy-influential scientists (PI-Sci). 
\textbf{(D)} Normalized rich-club coefficient \cite{colizza2006detecting} (tendency of top authors to inter-collaborate more than expected) of top policy-cited scientists' sub-network versus the fraction of scientists included. 
\textbf{(E)} Co-authorship networks of 300 most policy-cited scientists in four domains. Orange nodes: largest connected component; blue nodes: others. 
\textbf{(F)} Domain-level comparison. x-axis: size of largest connected component among top 300 scientists; y-axis: proportion of PI-Sci (scientists whose papers collectively account for 30\% of policy citations).}
    \label{fig:fig02}
\end{figure*}

\subsection{Policy-Influential Scientists} 

To assess how leading scientists shape international policy, we first assigned the 230,737 policy-cited papers to 23 policy domains by clustering the global paper-to-paper citation network (see \textit{Methods}; cluster details in Table S3). The resulting map (Figure~2A) spans the diverse spectrum of scientific knowledge referenced in IGO documents, with marked recent growth in areas such as \textit{Covid-19 Pandemic}, \textit{Circular Economy}, and \textit{Data Science \& AI} (Figure S2). These patterns highlight the heterogeneity of research underpinning contemporary policymaking.

Within every domain, policy citations are highly skewed toward a small subset of scholars, though the degree of concentration varies. The complementary cumulative distribution function (CCDF) of policy-cited papers per author follows a power-law-like form (Figure~2B), indicating that a handful of scientists attract disproportionately large citation shares. Domains such as \textit{Climate Modeling} and \textit{Marine Conservation} display relatively shallow slopes, indicative of strong concentration, whereas \textit{Data Science \& AI} shows a steeper slope and more diffuse influence distribution.

To characterize these influential contributors, we define Policy-Influential Scientists (PI-Sci) as authors whose weighted publications collectively account for 30\% of all policy-cited papers in a given domain, where each author receives $1/n$ credit for a paper with $n$ co-authors. Across domains, PI-Sci constitute fewer than 5\% of all policy-cited authors, underscoring their outsized role in bridging science and policy (Figure~2C). For example, in \textit{Climate Modeling}, just 0.62\% of scientists (258 individuals) generate 30\% of the policy-cited literature, a pattern echoed in \textit{Animal Diseases} and \textit{Marine Conservation}. By contrast, influence is more dispersed in \textit{Data Science \& AI} (3.5\%) and \textit{Environmental Toxicology} (2.0\%).

Collaboration among these top-cited scholars often forms dense, elite subnetworks. We quantified this tendency using the normalized rich-club coefficient\cite{colizza2006detecting}, where values $>1$ indicate that leading authors collaborate more frequently than expected by chance (Figure~2D). All 23 domains exceed this threshold, though direct comparison requires caution due to differences in network size and modularity.

Visualizing co-authorship among the 300 most policy-influential scientists (weighted by $1/n$) reveals striking structural differences (Figure~2E). \textit{Climate Modeling} is dominated by a single, tightly knit cluster; the \textit{Covid-19 Pandemic} domain fragments into several smaller groups; and \textit{Data Science \& AI} lacks any clear collaborative core, reflecting comparatively diffuse partnerships (networks for remaining domains shown in Figure S3).

Figure 2F illustrates a clear negative correlation between the proportion of top scientists achieving high policy citation rates and the size of their co-authorship networks, suggesting that stronger, more tightly connected networks enhance scientists' policy influence. For example, \textit{Climate Modeling} features greater dominance by influential authors within a dense collaborative network, reflecting strong internal cohesion and substantial policy impact. \textit{Environmental Toxicology} and \textit{Plastic Pollution} occupy intermediate positions, perhaps owing to the diverse chemical and technical specializations they encompass. At the opposite extreme, \textit{Data Science \& AI} lacks prominent policy-influential scientists and dense collaborative ties. Together, these structural contrasts illuminate how disciplinary network architecture shapes the translation of scientific consensus into policy.

\subsection{The Role of PI-Sci in Policymaking}

\begin{figure}[!h]
    \centering \includegraphics[width=0.75\linewidth]{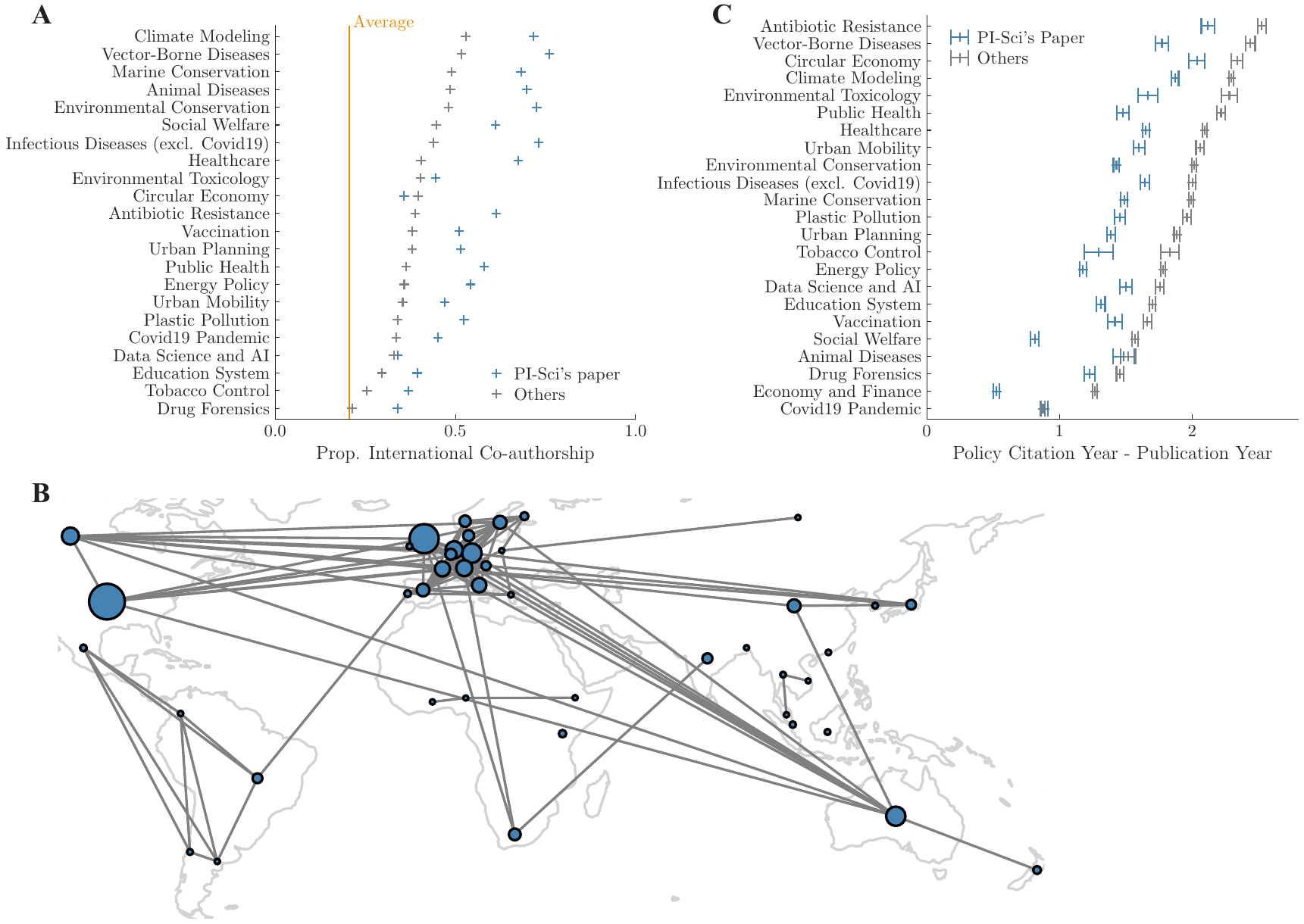}
    \caption{\textbf{Role of PI-Sci in policy-making} \textbf{(A)} International co-authorship across scientific domains. Blue crosses denote PI-Sci papers; gray crosses denote other policy-cited papers. The orange vertical line indicates the overall proportion of international co-authorship in Scopus-indexed publications (2014–2023). \textbf{(B)} Country-level co-authorship network among PI-Sci. Node size reflects the number of PI-Sci papers from each country. Edges connect country pairs whose co-authored PI-Sci papers exceed 6.5\% of their combined PI-Sci output. \textbf{(C)} Policy citation lag by research domain. Gray bars: average years between publication and policy citation for all papers. Blue bars are the same measure as PI-Sci papers. The dataset spans 2015-2023, yielding shorter lags than historical studies.}   
    \label{fig3new}
\end{figure}

To understand why a relatively small cohort of PI-Sci exerts such outsized influence, we examined the speed and structure of their engagement with policymakers. As shown in Figure~\ref{fig3new}A, PI-Sci papers display markedly higher rates of international co-authorship than other policy-cited papers across nearly every domain. The difference is most striking in fields such as \textit{Climate Modeling}, \textit{Vector-Borne Diseases}, and \textit{Environmental Conservation}, where PI-Sci papers routinely exceed both the average level of collaboration among policy-cited work and the overall Scopus baseline since 2014 (orange vertical line). In other words, policy-relevant science is already more international than the norm, and PI-Sci research pushes that boundary further.

Despite this global reach, collaboration remains unevenly distributed. Figure~\ref{fig3new}B reveals a dense nexus of co-authorship within the European Union. Here, node size represents the volume of PI-Sci output from each country, while edges link country pairs whose joint publications exceed 6.5\% of their combined PI-Sci corpus. This pattern reveals a worldwide yet regionally concentrated network that channels knowledge from key hubs into international policy forums.


Faster uptake is another characteristic of PI-Sci influence. In Figure~\ref{fig3new}C, the average lag between publication and first policy citation is one to two years for most scientists (gray bars) but is systematically shorter for PI-Sci across most domains (blue bars). The effect is particularly pronounced in \textit{Antibiotic Resistance}, \textit{Vector-Borne Diseases}, and \textit{Public Health}, suggesting that decision-makers preferentially draw on research produced by these highly connected scholars. The lone exception is Covid-19, where citation lags collapse for everyone—an expected outcome when global emergency accelerates the coevolution of science and policy\cite{yin2021coevolution}.


\subsection{Measuring the Policy Impact of PI-Sci in Climate Modeling}

\begin{figure}[!h]
    \centering
    \includegraphics[width=0.53\linewidth]{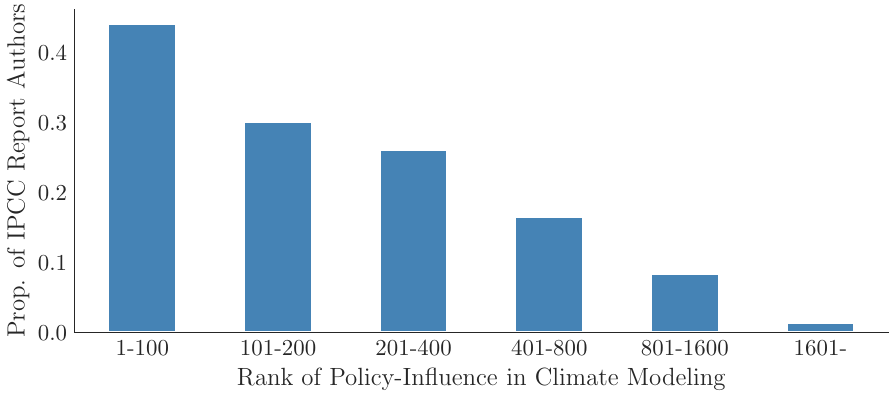}
    \caption{\textbf{Scientific advisory participation rates by research influence in Climate Modeling.} Proportion of climate modeling policy-cited scientists (n=41,594) serving as IPCC report authors, binned by policy-influence rank.}    
    \label{fig:PolicyInfluence}
\end{figure}

To disentangle factors driving the substantial and rapid policy impact of PI-Sci, we conducted a case study in \textit{climate modeling}, which features a well-established PI-Sci community. First, we assessed the overlap between PI-Sci and policymakers using IPCC authorship as a proxy for policymaker engagement. As shown in Figure~\ref{fig:PolicyInfluence}, over 40\% of the top 100 policy-influential climate-modeling scientists have served as IPCC authors. Among the 258 PI-Sci in this cluster, 97 (37.6\%) have contributed to IPCC reports, confirming that IPCC authorship strongly correlates with elevated policy influence.

Next, we investigated whether this heightened policy impact reflects scientists' proximity to policymakers or the intrinsic policy relevance of their research. Using cosine similarity-based semantic matching (see Methods), we paired each PI-Sci-authored paper with a non-PI-Sci paper of highly similar topical content. Despite near-identical subjects, PI-Sci papers received substantially more policy citations: in the matched sample, 50.1\% (1,228/2,450) of PI-Sci papers were cited in policy documents, compared to only 28.1\% (688/2,450) for non-PI-Sci papers. These results suggest that enhanced citation rates of PI-Sci papers may arise from close relationships between PI-Sci and policymakers at IGOs.

Our analysis reveals a clear division in the types of scientific terms emphasized by Policy-Influential Scientists (PI-Sci) compared to other scientists, highlighting complementary roles in informing policy. To quantify this distinction, we used the \textit{PI-Sci Papers Rate} (PPR), which represents the proportion of each term's total occurrences in climate modeling papers authored by at least one PI-Sci. Terms with the lowest PPR values (PPR = 0) highlight foundational observational efforts such as methane tracking and biodiversity monitoring. Intermediate PPR ranges emphasize regional impacts and socioeconomic dimensions (e.g., catastrophic wildfires). In contrast, the highest PPR bins are dominated by advanced modeling tools and scenario simulations—such as Shared Socioeconomic Pathways (SSPs) and CMIP6 ensemble experiments—that directly inform strategic policy decisions. At PPR = 1, highly specialized phrases like "mixed-layer ocean model" or "experiment perfect predictor" prevail, illustrating PI-Sci's pivotal role in supplying sophisticated modeling frameworks on which global climate policy increasingly depends.

\begin{table}[!h]
\centering
\begin{tabular}{m{3.5cm}m{8.45cm}}
\toprule
\centering \textbf{PPR} & \multicolumn{1}{c}{\textbf{Representative Terms}} \\ \midrule
\centering \textbf{$\mathrm{PPR}=0$} & change climate towards, (prd) delta, (prd) delta river, methane report, earth engine, satellite sentinel-2, biodiversity circumpolar monitoring, 220 du, biodiversity freshwater, tropomi using, irrigation requirement water, groundwater withdrawal, beijing climate, earth engine google, land use/cover
 \\ \hline
\centering \textbf{$0<\mathrm{PPR}<0.684$} & ecological social, co2 global, know phenomenon, event respectively, difference highlight, catastrophic wildfire, lead turn, two urban, contribution runoff, data tropomi, gosat satellite, atmospheric layer, basin coal, assumption simplify, average climatological \\ \hline
\centering \textbf{$0.684 \leq \mathrm{PPR} < 1$} & measurement tccon, scenario ssp5-8.5, cfc-11 global, cmip6 multi-model, america northern, ssp2-4.5 ssp3-7.0, resolution standard, elements tip, associate driver, cmip6 ensemble, (mpi-ge) ensemble grand, (mpi-ge) ensemble, assessment sixth, ssp1-2.6 ssp2-4.5, assessment report sixth \\ \hline
\centering \textbf{$\mathrm{PPR}=1$} & perfect predictor value, experiment predictor, mixed-layer model ocean, perfect value, cmip6 ensemble multi-model, experiment perfect predictor
 \\ \bottomrule
\end{tabular}
\caption{Representative terms from climate-modeling papers (2018-2023) by \textit{PI-Sci Papers Rate} (PPR). Bi-gram and tri-gram terms are extracted (see Methods). PPR indicates the ratio of term frequency in PI-Sci papers (31.6\% of total) to frequency in non-PI-Sci papers.}
\label{tab:KeyPhrase_table}
\end{table}

In sum, our analysis demonstrates that policy-influential scientists in climate modeling exert a disproportionate impact on policy not simply because they study salient topics, but because their professional stature—reflected in IPCC authorship and other forms of visibility—amplifies the uptake of their work. Even after controlling for topical similarity, PI-Sci papers receive markedly more policy citations, and text mining reveals a division of labor in which PI-Sci contribute high-level modeling frameworks while non-PI-Sci supply crucial observational and problem-identification research. Together, these findings highlight the importance of both scientific content and scientific prominence in shaping how climate knowledge informs policy.

\begin{figure}[!h]
    \centering \includegraphics[width=0.5\linewidth]{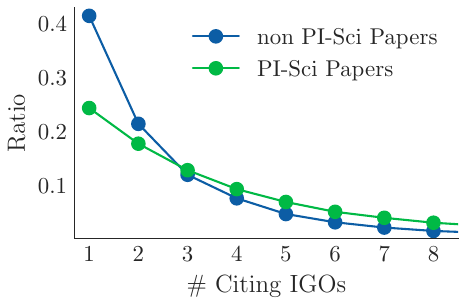}
    \caption{\textbf{Distribution of number of IGOs citing papers.} Comparison between papers authored by Policy-Influential Scientists (PI-Sci, green) and other papers (non-PI-Sci, blue).}
    \label{fig8}
\end{figure}

\subsection{Synchronization in IGOs' Policy Citations}

Policy-influential scientists (PI-Sci) publish research that spreads through a broader range of intergovernmental organizations (IGOs) than work authored by other scientists. As illustrated in Figure~\ref{fig8}, PI-Sci papers are cited less often by only a single IGO and more often by two or more IGOs relative to non-PI-Sci papers. On average, a PI-Sci paper is cited by 1.73 times as many IGOs as a non-PI-Sci paper, a pattern that holds across all policy domains. This multi-IGO "source-node" position enables PI-Sci research to seed coordinated uptake of science across institutional boundaries.

\begin{figure}[!h]
    \centering \includegraphics[width=0.8\linewidth]{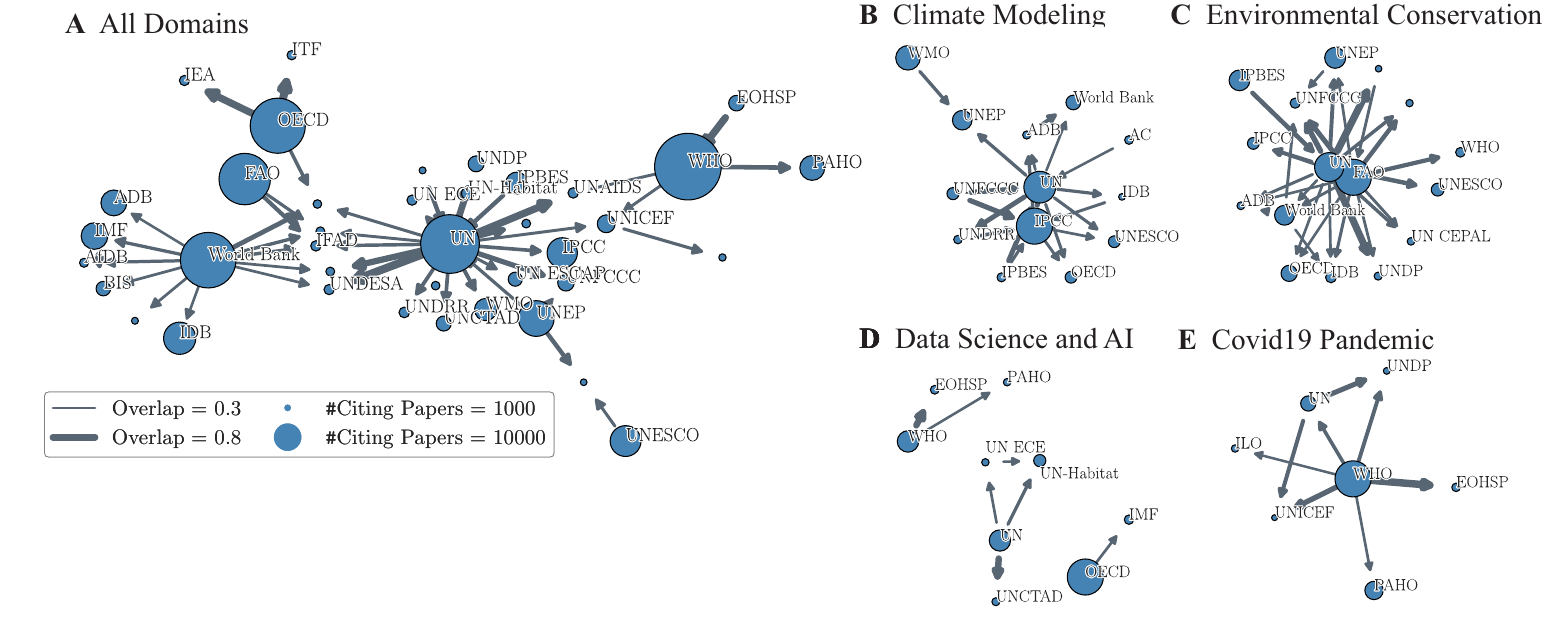}
\caption{\textbf{Co-citation networks of IGOs.} \textbf{(A)} Aggregate network across all domains. Node size proportional to citing papers. Edge width indicates citation-set overlap (co-cited papers divided by a smaller citation set); only edges with overlap $>0.3$ are shown. Arrows point from earlier-citing to later-citing IGO. \textbf{(B-E)} Domain-specific networks.}    
\label{fig6}
\end{figure}

The aggregate co-citation network (Figure~\ref{fig6}A) reveals a pronounced temporal hierarchy in the flow of scientific information among IGOs. Arrows point from organizations that cite papers earlier to those that cite them later, showing that large, broadly mandated institutions—most notably the United Nations (UN), World Bank, and World Health Organization (WHO)—consistently integrate new evidence first. Smaller, more specialized bodies enter the citation stream later, suggesting that major multilateral organizations act as primary gateways through which scientific knowledge diffuses to the wider policy ecosystem.

Domain-specific subnetworks clarify how this cascade varies by field (Figure~\ref{fig6}B-E). In the mature area of \textit{Climate Modeling}, a dense core is anchored by IPCC and UN (Figure~\ref{fig6}B), whose citation profile overlaps strongly with that of the UN (overlap = 0.59). Similarly, there is a dense core(UN, FAO) and disconnection in \textit{Environmental Conservation}(Figure~\ref{fig6}C). A similarly cohesive structure emerges for the \textit{Covid-19 Pandemic}, where the WHO synchronizes citations across numerous public-health-oriented IGOs. By contrast, \textit{Data Science and AI} and  \textit{Covid-19 Pandmic} display a fragmented topology with no single dominant hub, reflecting the field's relative nascency(Figure~\ref{fig6}D, E).

Collectively, these patterns indicate that research policy influence is amplified when it diffuses hierarchically and synchronously through leading IGOs, and that PI-Sci research benefits most from this amplification.

\section{Discussion}  

Our analysis reveals that a remarkably small cohort of policy-influential scientists (PI-Sci) acts as a conduit through which research is translated into intergovernmental policy. Echoing "honest broker" and boundary-spanner concepts\cite{cash2003knowledge,gluckman2021brokerage}, these scientists attract disproportionate shares of IGO citations, receive them sooner, and appear synchronously across multiple organizations. This pattern builds upon prior research linking scholarly impact to policy uptake\cite{yin2021coevolution}, which showed that policy-cited scientific papers tend to be highly cited within academia, with underlying human networks facilitating this connection. Our findings show that PI-Sci papers receive multiple IGO citations, are cited more rapidly, spread synchronously across organizations, and remarkably, PI-Sci themselves often serve within IGO advisory bodies. This reveals a co-evolution of policymakers and policy-influential scientists, where sustained interactions and overlapping memberships create a dynamic feedback loop between knowledge production and policy formation. The result is a pronounced Matthew effect\cite{merton1968matthew}: once scientists gain visibility in policy circles, cumulative advantages accelerate further recognition, mirroring well-documented dynamics in academic citation networks\cite{reardon2021elite}.

The structure of underlying scientific fields conditions how this advantage forms. In highly organized domains such as \textit{Climate Modeling}, a dense rich-club network channels citations toward a handful of IPCC-connected scholars; fewer than 1\% of authors account for 30\% of policy-cited output. Similar elite clusters have been documented in epistemic-community studies\cite{haas1992introduction} and bibliometric work on "rich clubs" in science\cite{colizza2006detecting}. By contrast, emerging or fragmented areas—e.g., AI governance\cite{cihon2020fragmentation}—lack a single coordinating hub and therefore show more even, albeit weaker, policy presence. These field-level differences suggest two pathways to evidence uptake: centralized consensus, which delivers clear, rapidly actionable advice (beneficial for climate mitigation), and distributed pluralism, which may surface diverse perspectives but at the cost of slower synthesis.

Variation across IGOs further shapes who becomes influential. Organizations with formal assessment machinery (IPCC, IPBES, WHO) institutionalize science advice and systematically elevate interface scientists through authorship roles\cite{vasileiadou2011exploring}. Club-like bodies such as the OECD or development banks, whose membership and missions vary, draw more selectively on external expertise, potentially reinforcing regional or disciplinary biases\cite{ho2011participates}. Policy-network theory predicts that actors occupying high betweenness positions—here, PI-Sci embedded in multiple advisory arenas—control information flows. The synchronicity in citation across organizations supports this brokerage model: the same names and papers recur as IGOs cite one another's evidence bases.

These dynamics have governance implications. Concentrating advice in tight networks can expedite decision-making yet risks epistemic lock-in, under-representation of scholars from the Global South, and blind spots for emerging issues. Recent calls for inclusive science advice—from the G20 Chief Science Advisers' Roundtable\cite{basha2023cohesive} to the UN's Governing AI for Humanity report\cite{AIAdviso37:online}—underline the need to widen participation without sacrificing rigor. Existing efforts include the International Network for Government Science Advice (INGSA), which develops the science-policy interface in developing countries\cite{cowen2020effective}. Mechanisms such as rotating committee membership, capacity-building for early-career and under-represented scientists, and open evidence portals could mitigate excessive path dependence while retaining the benefits of expert continuity. Particularly in fields such as AI, establishing governance systems that keep pace with rapid technological progress while accommodating diverse perspectives is crucial\cite{cosens2021governing, oecd2022harnessing}. As private-sector scientists play increasingly important roles in technological development, we must promote public-private collaboration\cite{alliance2024briefing} while maintaining neutrality and objectivity in science advice for international policymaking\cite{havstad2017neutrality}.

Several limitations qualify our conclusions. Bibliometric indicators capture only cited knowledge, omitting informal exchanges, grey literature, and the politics of advisory selection. Although Overton offers comparatively broader coverage than other databases\cite{maleki2024policy,WhatisOv15:online}, coverage biases can still obscure the influence of research from regions with limited documentation. Future work should integrate qualitative mapping of advisory bodies, topic modeling of policy text, and longitudinal tracking of how policy feedback reshapes research agendas (e.g., post-SDG funding shifts). Comparative network analyses across broader ranges of IGOs could identify institutional features—size, mandate, stakeholder diversity, geographical reach—that foster balanced and timely science advice.

In sum, interface scientists are pivotal nodes in the global knowledge-governance web. Their influence is produced jointly by field-specific collaboration patterns and advisory architectures of IGOs. Making these interfaces more transparent and inclusive is essential for robust, equitable, and agile evidence-based policymaking.

\section{Methods}

\subsection{Dataset}
We used two primary data sources: Overton and Scopus. Overton is a comprehensive database of policy documents and their citations to scientific papers, whereas Scopus was chosen for its extensive coverage and accurate author identification \cite{baas2020scopus}. Our analysis focused on policy-cited papers from intergovernmental organizations (IGOs) and was restricted to papers published since 2015 to capture recent trends in science-policy interactions.

From Overton, we initially retrieved 305,635 papers published since 2014 that were cited by IGO policy documents. We then matched these Overton records with Scopus entries using DOIs. This matching allowed us to identify and link papers in Scopus that had been cited by policy documents in Overton. After this process, the final dataset comprised 230,737 papers that could be definitively linked between Overton and Scopus.

\subsection{Measuring Distance to Policy}
Following the concept of paper patent distance\cite{ahmadpoor2017dual}, we defined each paper's distance to policy as the shortest undirected path connecting it to a policy document in a combined citation network of both scientific papers and IGO policy documents. In this network, each node represents either a paper or a policy document, and edges represent citation relationships that can be traversed in both directions. Papers directly cited by a policy document have a distance of 1. For any other paper, the distance is 1 plus the shortest path to any distance 1 paper, regardless of the citation direction. For instance, a paper with distance 2 may either cite or be cited by a distance 1 paper but is not directly cited by policy.

\subsection{Estimating the Number of Policy-Cited Papers for Each Country}

To estimate how many papers from each country are cited in IGO policy documents, we developed four methods. Each method incrementally uses more detailed information to enhance the accuracy of the estimates:

\begin{itemize}

\item \textbf{Method 1 (N (Total number of papers))}:
This simplest method only considers the total number of papers published by each country. All papers from a country are treated equally, without distinguishing among different characteristics.

\item \textbf{Method 2 (N + J (Journal categories))}:
This method improves accuracy by considering which journal papers were published. We group papers by journal and estimate the probability of being cited in policy documents based on how often papers from each journal have historically been cited.

\item \textbf{Method 3 (N + J + R ( Reference lists))}:
This method further refines the estimates by looking at each paper's reference list. Papers are grouped according to how many of their references have previously been cited in policy documents. To handle wide ranges, we group these counts into logarithmic bins (e.g., 0, 1, 2–3, 4–7 references, and so forth).

\item \textbf{Method 4 (N + J + R + A (Authors’ prior policy-cited papers ))}:
The most comprehensive method also considers the past publication history of the authors. Papers are grouped based on the maximum number of policy-cited papers previously published by any of their authors, again using logarithmic bins.

\end{itemize}

For each of these methods, we first used papers published \emph{before} our estimation period to create groups based on relevant features (journal, references, authors’ prior policy-cited papers ). Within each group, we calculated the historical probability that a paper is cited by policy documents. Specifically, for a group $g$ containing $n_g$ total papers, of which $m_g$ were cited in policy documents, the probability is computed as $
P(\text{policy-cited} \mid g) = \frac{m_g}{n_g}
$.

To estimate how many papers from each country are cited in policy documents during the estimation period, we sum these probabilities across all papers from the country. Formally, if paper $i$ belongs to country $x$ and is part of group $g_i$, its probability of being cited in policy documents is $P(\text{policy-cited} \mid g_i)$. Therefore, the estimated total number of policy-cited papers for country $x$ is calculated as $
\sum_{i \in C_x} P(\text{policy-cited} \mid g_i),
$, where $C_x$ denotes the set of all papers affiliated with country $x$.

\subsection{Grouping Policy-Cited Papers by Research Domains}

To identify major research themes among the papers cited in policy documents, we first constructed a citation network of these papers. We then used the Leiden clustering method \cite{traag2019louvain}, which identifies groups (clusters) of papers that frequently cite each other, thereby highlighting distinct research domains.

After identifying these clusters, we characterized each domain by analyzing the abstracts of the papers within it. Specifically, we calculated TF-IDF scores—indicators of the most important and distinctive terms—for each cluster. Finally, we used GPT-4 to summarize these terms into clear and concise labels, describing the main theme of each research domain.

\subsection{Identifying Policy-Influential Scientists (PI-Sci) in Each Research Domain}

Within each research domain, we identified scientists whose research significantly influenced policy—termed Policy-Influential Scientists (PI-Sci)—using the following procedure: First, we considered each policy-cited paper within a domain. If a paper had $n$ authors, each author was given an equal fractional contribution of $1/n$. We then calculated each scientist's total contribution by summing these fractional values across all their policy-cited papers within the domain. Next, we ranked the scientists in descending order based on these summed contributions. Finally, we selected the top scientists whose combined contributions accounted for 30\% of all policy-cited papers within that domain. This method ensures that the identified PI-Sci reflect the influential scientists within their respective thematic areas.

\subsection{Matching Semantically Similar PI-Sci and Non-PI-Sci's Papers}
To assess how PI-Sci authorship relates to policy impact in climate change research, we used Elsevier’s International Centre for the Study of Research (ICSR) dataset, which contains metadata on over 90 million scholarly papers indexed in the Scopus database. The ICSR dataset includes Elsevier’s United Nations Sustainable Development Goals (SDG) research classifications, which categorize research papers as contributing to or addressing one of the SDGs\cite{bedardvallee_sdg_2023}. Using the SDG categories, we identified 338,250 research papers related to SDG 13: Climate Action published after 2016, of which 5,167 had at least one PI-Sci author.

For each PI-Sci-authored paper in the dataset, we searched for semantically similar papers across 20 million Scopus papers published after 2016 and before May 2023. The search used miniLM\cite{wang2020minilm} generated abstract embeddings and elastic KNN\cite{knn} to find the top 10 nearest neighbors with a cosine similarity score above 0.8, a threshold validated manually in previous research\cite{mahfouz5197403uncovering}. The matches were filtered to include pairs where both papers were published no more than one year apart, had entirely different author teams, and where one paper had at least one PI-Sci co-author while the other had no PI-Sci authors at all. This resulted in a final dataset of 2,450 pairs, with some PI-Sci authored papers having multiple matches which we treat as independent pairs.

\subsection{Author Matching between IPCC Members and PI-Sci} We compiled author lists from 12 IPCC assessments and special reports published between 2011 and 2023 (detailed in Table S4). Names were extracted from the PDF appendices, converted to lowercase in a standardized “surname, given name” format, and de-duplicated, resulting in a list of 1,639 unique authors.

To identify overlaps with the 258 policy-influential scientists (PI-Sci) in the \textit{Climate Modeling} domain, we applied fuzzy string matching using the token\_set\_ratio function from the FuzzyWuzzy library. For each IPCC author, we identified the best-matching name in the PI-Sci list based on the token set ratio score. Matches with a score above 83 were considered valid; this threshold was selected to allow for minor name variations while minimizing false positives. In total, 97 PI-Sci were identified as authors on at least one of the selected IPCC reports.

\subsection{Identifying Emerging Terms in Climate Modeling Papers}
We analyzed recent trends in climate modeling research by identifying new scientific terms introduced from 2018 onward. Although our dataset starts from 2015, we began the analysis from 2018 to allow for a sufficient buffer period to accurately capture emerging terminology.
We extracted two- and three-word phrases (bi-grams and tri-grams) from abstracts of climate modeling papers published from 2018 onwards. To standardize these terms, we applied lemmatization, which converts words to their base forms. Additionally, we recorded the first year each term appeared in our dataset.
We categorized papers into two groups based on authorship: "PI-Sci Papers," which included at least one coauthor identified as a Policy-Influential Scientist (PI-Sci), and "non-PI-Sci Papers," which had no PI-Sci coauthors. Of the 6,841 climate modeling papers analyzed, 2,162 (31.6\%) were PI-Sci Papers, while 4,679 (68.4\%) were non-PI-Sci Papers.

For each phrase, we calculated the proportion of occurrences in PI-Sci Papers relative to its total occurrences within all climate modeling papers. This proportion was labeled the \textit{PI-Sci Papers Rate} (PPR). Based on PPR values, phrases were classified into four distinct categories:
\begin{itemize}
 \item \textbf{[0]}: Terms appearing exclusively in non-PI-Sci papers.
 \item \textbf{(0, 0.684)}: Terms appearing proportionally less frequently in PI-Sci papers than the overall proportion of PI-Sci papers.
 \item \textbf{[0.684, 1)}: Terms appearing proportionally more frequently in PI-Sci papers than the overall proportion of PI-Sci papers.
 \item \textbf{[1]}: Terms appearing exclusively in PI-Sci papers.
 \end{itemize}
This classification helped identify terms that are uniquely emphasized or potentially overlooked by PI-Sci in climate modeling research domain.

\bibliographystyle{unsrt}
\bibliography{main}

\begin{thebibliography}{10}

\bibitem{cash2003knowledge}
David~W Cash, William~C Clark, Frank Alcock, Nancy~M Dickson, Noelle Eckley,
  David~H Guston, Jill J{\"a}ger, and Ronald~B Mitchell.
\newblock Knowledge systems for sustainable development.
\newblock {\em Proceedings of the national academy of sciences},
  100(14):8086--8091, 2003.

\bibitem{contandriopoulos2010knowledge}
Damien Contandriopoulos, Marc Lemire, Jean-Louis Denis, and {\'E}mile Tremblay.
\newblock Knowledge exchange processes in organizations and policy arenas: a
  narrative systematic review of the literature.
\newblock {\em The Milbank Quarterly}, 88(4):444--483, 2010.

\bibitem{gerber2023bridging}
Leah~R Gerber.
\newblock Bridging the gap between science and policy for a sustainable future.
\newblock {\em Nature Water}, 1(10):824--824, 2023.

\bibitem{haas1992introduction}
Peter~M Haas.
\newblock Introduction: epistemic communities and international policy
  coordination.
\newblock {\em International organization}, 46(1):1--35, 1992.

\bibitem{yin2021coevolution}
Yian Yin, Jian Gao, Benjamin~F. Jones, and Dashun Wang.
\newblock Coevolution of policy and science during the pandemic.
\newblock {\em Science}, 371(6525):128--130, 2021.

\bibitem{gupta2010history}
Joyeeta Gupta.
\newblock A history of international climate change policy.
\newblock {\em Wiley Interdisciplinary Reviews: Climate Change}, 1(5):636--653,
  2010.

\bibitem{gluckman2021brokerage}
Peter~D Gluckman, Anne Bardsley, and Matthias Kaiser.
\newblock Brokerage at the science--policy interface: from conceptual framework
  to practical guidance.
\newblock {\em Humanities and Social Sciences Communications}, 8(1):1--10,
  2021.

\bibitem{dwivedi2024real}
Yogesh~K Dwivedi, Anand Jeyaraj, Laurie Hughes, Gareth~H Davies, Manju Ahuja,
  Mousa~Ahmed Albashrawi, Adil~S Al-Busaidi, Salah Al-Sharhan, Khalid~Ibrahim
  Al-Sulaiti, Levent Altinay, et~al.
\newblock “real impact”: Challenges and opportunities in bridging the gap
  between research and practice--making a difference in industry, policy, and
  society, 2024.

\bibitem{akerlof2022global}
KL~Akerlof, Alessandro Allegra, Selena Nelson, Cameryn Gonnella, Carla
  Washbourne, and Chris Tyler.
\newblock Global perspectives on scientists’ roles in legislative
  policymaking.
\newblock {\em Policy Sciences}, 55(2):351--367, 2022.

\bibitem{ahmadpoor2017dual}
Mohammad Ahmadpoor and Benjamin~F Jones.
\newblock The dual frontier: Patented inventions and prior scientific advance.
\newblock {\em Science}, 357(6351):583--587, 2017.

\bibitem{fraiberger2018quantifying}
Samuel~P Fraiberger, Roberta Sinatra, Magnus Resch, Christoph Riedl, and
  Albert-L{\'a}szl{\'o} Barab{\'a}si.
\newblock Quantifying reputation and success in art.
\newblock {\em Science}, 362(6416):825--829, 2018.

\bibitem{szomszor2022overton}
Martin Szomszor and Euan Adie.
\newblock Overton: A bibliometric database of policy document citations.
\newblock {\em Quantitative science studies}, 3(3):624--650, 2022.

\bibitem{bornmann2022relevant}
Lutz Bornmann, Robin Haunschild, Kevin Boyack, Werner Marx, and Jan~C Minx.
\newblock How relevant is climate change research for climate change policy? an
  empirical analysis based on overton data.
\newblock {\em PloS one}, 17(9):e0274693, 2022.

\bibitem{dorta2025kind}
Pablo Dorta-Gonz{\'a}lez.
\newblock Which kind of research papers influence policymaking.
\newblock {\em arXiv preprint arXiv:2504.18889}, 2025.

\bibitem{dorta2024societal}
Pablo Dorta-Gonz{\'a}lez, Alejandro Rodr{\'\i}guez-Caro, and Mar{\'\i}a~Isabel
  Dorta-Gonz{\'a}lez.
\newblock Societal and scientific impact of policy research: A large-scale
  empirical study of some explanatory factors using altmetric and overton.
\newblock {\em Journal of Informetrics}, 18(3):101530, 2024.

\bibitem{anderson2020case}
P~Sage Anderson, Aubrey~R Odom, Hunter~M Gray, Jordan~B Jones, William~F
  Christensen, Todd Hollingshead, Joseph~G Hadfield, Alyssa Evans-Pickett,
  Megan Frost, Christopher Wilson, et~al.
\newblock A case study exploring associations between popular media attention
  of scientific research and scientific citations.
\newblock {\em PLoS One}, 15(7):e0234912, 2020.

\bibitem{tomokiyo2025researchers}
Yuta Tomokiyo, Keita Nishimoto, Kimitaka Asatani, and Ichiro Sakata.
\newblock Do researchers benefit career-wise from involvement in international
  policy guideline development?
\newblock {\em arXiv preprint arXiv:2503.22584}, 2025.

\bibitem{furnas2025partisan}
Alexander~C Furnas, Timothy~M LaPira, and Dashun Wang.
\newblock Partisan disparities in the use of science in policy.
\newblock {\em Science}, 388(6745):362--367, 2025.

\bibitem{van2007rationale}
Sybille Van~den Hove.
\newblock A rationale for science--policy interfaces.
\newblock {\em Futures}, 39(7):807--826, 2007.

\bibitem{oliver2014new}
Kathryn Oliver, Theo Lorenc, and Simon Innv{\ae}r.
\newblock New directions in evidence-based policy research: a critical analysis
  of the literature.
\newblock {\em Health research policy and systems}, 12:1--11, 2014.

\bibitem{parkhurst2017politics}
Justin Parkhurst.
\newblock {\em The politics of evidence: from evidence-based policy to the good
  governance of evidence}.
\newblock Taylor \& Francis, 2017.

\bibitem{merton1968matthew}
Robert~K Merton.
\newblock The matthew effect in science: The reward and communication systems
  of science are considered.
\newblock {\em Science}, 159(3810):56--63, 1968.

\bibitem{jacomy2014forceatlas2}
Mathieu Jacomy, Tommaso Venturini, Sebastien Heymann, and Mathieu Bastian.
\newblock Forceatlas2, a continuous graph layout algorithm for handy network
  visualization designed for the gephi software.
\newblock {\em PloS one}, 9(6):e98679, 2014.

\bibitem{colizza2006detecting}
Vittoria Colizza, Alessandro Flammini, M~Angeles Serrano, and Alessandro
  Vespignani.
\newblock Detecting rich-club ordering in complex networks.
\newblock {\em Nature physics}, 2(2):110--115, 2006.

\bibitem{reardon2021elite}
Sara Reardon.
\newblock Elite’researchers dominate citation space.
\newblock {\em Nature}, 591(7849):333--334, 2021.

\bibitem{cihon2020fragmentation}
Peter Cihon, Matthijs~M Maas, and Luke Kemp.
\newblock Fragmentation and the future: Investigating architectures for
  international ai governance.
\newblock {\em Global Policy}, 11(5):545--556, 2020.

\bibitem{vasileiadou2011exploring}
Eleftheria Vasileiadou, Gaston Heimeriks, and Arthur~C Petersen.
\newblock Exploring the impact of the ipcc assessment reports on science.
\newblock {\em Environmental Science \& Policy}, 14(8):1052--1061, 2011.

\bibitem{ho2011participates}
Claudia Ho-Lem, Hisham Zerriffi, and Milind Kandlikar.
\newblock Who participates in the intergovernmental panel on climate change and
  why: A quantitative assessment of the national representation of authors in
  the intergovernmental panel on climate change.
\newblock {\em Global Environmental Change}, 21(4):1308--1317, 2011.

\bibitem{basha2023cohesive}
B~Chagun Basha and Parvinder Maini.
\newblock For cohesive solutions, scientific advice must be accessible and
  widely drawn.
\newblock {\em Nature India}, 2023.

\bibitem{AIAdviso37:online}
United Nations.
\newblock Ai advisory body | united nations, Sep 2024.
\newblock [Online; accessed 2025-05-17].

\bibitem{cowen2020effective}
Lara Cowen et~al.
\newblock Effective science advice to governments in the developing world:
  final technical report.
\newblock 2020.

\bibitem{cosens2021governing}
Barbara Cosens, JB~Ruhl, Niko Soininen, Lance Gunderson, Antti Belinskij,
  Thorsten Blenckner, Alejandro~E Camacho, Brian~C Chaffin, Robin~Kundis Craig,
  Holly Doremus, et~al.
\newblock Governing complexity: Integrating science, governance, and law to
  manage accelerating change in the globalized commons.
\newblock {\em Proceedings of the National Academy of Sciences},
  118(36):e2102798118, 2021.

\bibitem{oecd2022harnessing}
OECD.
\newblock Harnessing the power of ai and emerging technologies: Background
  paper for the cdep ministerial meeting, 2022.

\bibitem{alliance2024briefing}
John~Granger Paul~Daugherty, Jeremy~Jurgens and Cathy Li.
\newblock Ai governance alliance: Briefing paper series, 2024.

\bibitem{havstad2017neutrality}
Joyce~C Havstad and Matthew~J Brown.
\newblock Neutrality, relevance, prescription, and the ipcc.
\newblock {\em Public Affairs Quarterly}, 31(4):303--324, 2017.

\bibitem{maleki2024policy}
Ashraf Maleki and Kim Holmberg.
\newblock Policy citations tracked by overton. io versus altmetric. com: Case
  study of finnish research organizations in social sciences.
\newblock {\em Informaatiotutkimus}, 43(3-4):4--28, 2024.

\bibitem{WhatisOv15:online}
What is overton's coverage and how does it compare to other systems? - overton
  knowledge base.
\newblock [Online; accessed 2025-05-19].

\bibitem{baas2020scopus}
Jeroen Baas, Michiel Schotten, Andrew Plume, Gr{\'e}goire C{\^o}t{\'e}, and
  Reza Karimi.
\newblock Scopus as a curated, high-quality bibliometric data source for
  academic research in quantitative science studies.
\newblock {\em Quantitative science studies}, 1(1):377--386, 2020.

\bibitem{traag2019louvain}
Vincent~A Traag, Ludo Waltman, and Nees~Jan Van~Eck.
\newblock From louvain to leiden: guaranteeing well-connected communities.
\newblock {\em Scientific reports}, 9(1):1--12, 2019.

\bibitem{bedardvallee_sdg_2023}
Alexandre Bedard-Vallee, Chris James, and Guillaume Roberge.
\newblock Elsevier 2023 sustainable development goals (sdgs) mapping, 2023.
\newblock Accessed: 2025-05-17.

\bibitem{wang2020minilm}
Wenhui Wang, Furu Wei, Li~Dong, Hangbo Bao, Nan Yang, and Ming Zhou.
\newblock Minilm: Deep self-attention distillation for task-agnostic
  compression of pre-trained transformers.
\newblock {\em Advances in Neural Information Processing Systems},
  33:5776--5788, 2020.

\bibitem{knn}
Thomas Cover and Peter Hart.
\newblock Nearest neighbor pattern classification.
\newblock {\em IEEE transactions on information theory}, 13(1):21--27, 1967.

\bibitem{mahfouz5197403uncovering}
Basil Mahfouz, Licia Capra, and Geoff Mulgan.
\newblock Uncovering drivers of science use in climate policy with pretrained
  language models.
\newblock {\em Available at SSRN 5197403}, 2025.

\end{thebibliography}

\clearpage

\part*{Supplementary Information}
\setcounter{section}{0}
\setcounter{table}{0}
\setcounter{figure}{0}

\vspace{-1cm}

\begin{table*}[ht]
\vspace{-1cm}
\caption*{\textbf{Table S1.  Country-level comparison of observed versus expected policy citations across four models
} For the 40 most prolific producer countries, each column shows observed papers cited in policy documents (Act), model's expected count (Est), and their ratio (Act/Est). When papers have international co-authors, the number of policy-cited papers are assigned to countries based on the share of the authors' affiliation countries. Models cumulatively add: Method 1 - publication volume (N); Method 2 - journal identity (J); Method 3 - policy-cited references per paper (R); Method 4 - authors' maximum past policy-citation count (A). Ratios $> 1$ indicate overrepresentation in policy citations; ratios $< 1$ indicate underrepresentation. Progressive factor incorporation reduces disparities: China improves from 0.37 to 0.84, UK decreases from 2.47 to 1.07, demonstrating that journal choice and author experience explain initial variation. In Method 4, 26 of the 40 countries fall within 0.9--1.1, though some East Asian and Eastern European countries remain underrepresented.
}
\small
\centering
\begin{tabular}{lccc|c|c|c}
\toprule
\textbf{Country} & \multicolumn{3}{c|}{\textbf{Estimated by number of papers(N)}} & \parbox{2cm}{\textbf{N + Journal(J)}} & \parbox{2cm}{\textbf{N + J + Number of policy-cited refs(R) }} & \parbox{2cm}{\textbf{N + J + R + Author's past policy-citation} } \\
\cmidrule(lr){2-7} 
 & \textbf{Act} & \textbf{Est} & \textbf{Act/Est} &   \textbf{Act/Est} & \textbf{Act/Est} & \textbf{Act/Est} \\
\midrule
chn & 2376.18 & 6502.84 & 0.37 & 0.61 & 0.73 & 0.84 \\
usa & 6092.14 & 4422.48 & 1.38 & 0.97 & 0.96 & 0.97 \\
ind & 603.05 & 1504.93 & 0.40 & 0.95 & 1.02 & 1.04 \\
gbr & 2768.90 & 1121.08 & 2.47 & 1.27 & 1.16 & 1.07 \\
deu & 1221.63 & 1024.30 & 1.19 & 1.02 & 0.96 & 0.96 \\
jpn & 387.83 & 872.09 & 0.44 & 0.67 & 0.81 & 0.86 \\
ita & 943.81 & 833.88 & 1.13 & 1.11 & 1.10 & 1.09 \\
rus & 108.38 & 898.03 & 0.12 & 0.75 & 0.86 & 0.87 \\
fra & 698.24 & 614.23 & 1.14 & 1.02 & 1.01 & 1.00 \\
can & 993.60 & 618.82 & 1.61 & 1.00 & 0.96 & 0.96 \\
esp & 720.68 & 640.03 & 1.13 & 0.92 & 0.92 & 0.95 \\
kor & 287.18 & 613.80 & 0.47 & 0.59 & 0.74 & 0.82 \\
aus & 1242.14 & 581.23 & 2.14 & 1.16 & 1.04 & 0.98 \\
bra & 572.30 & 619.33 & 0.92 & 0.98 & 0.99 & 1.01 \\
irn & 230.08 & 479.87 & 0.48 & 0.75 & 0.87 & 0.92 \\
tur & 172.27 & 377.83 & 0.46 & 0.83 & 0.91 & 0.99 \\
pol & 170.05 & 361.49 & 0.47 & 0.69 & 0.79 & 0.89 \\
nld & 820.86 & 309.88 & 2.65 & 1.27 & 1.10 & 1.01 \\
idn & 152.14 & 366.42 & 0.42 & 1.32 & 1.23 & 1.18 \\
twn & 84.23 & 242.11 & 0.35 & 0.38 & 0.48 & 0.58 \\
che & 564.69 & 219.96 & 2.57 & 1.31 & 1.17 & 1.04 \\
mys & 119.01 & 225.24 & 0.53 & 0.94 & 0.95 & 0.98 \\
swe & 500.60 & 201.40 & 2.49 & 1.23 & 1.08 & 1.04 \\
bel & 385.63 & 167.18 & 2.31 & 1.39 & 1.16 & 1.06 \\
prt & 196.74 & 174.11 & 1.13 & 1.05 & 1.01 & 1.01 \\
mex & 180.91 & 181.60 & 1.00 & 1.06 & 1.09 & 1.07 \\
zaf & 423.33 & 164.16 & 2.58 & 1.45 & 1.26 & 1.13 \\
egy & 111.35 & 163.60 & 0.68 & 1.17 & 1.27 & 1.20 \\
dnk & 289.55 & 141.29 & 2.05 & 1.14 & 1.08 & 1.02 \\
cze & 64.55 & 140.42 & 0.46 & 0.69 & 0.76 & 0.82 \\
sau & 138.01 & 155.01 & 0.89 & 1.20 & 1.27 & 1.24 \\
pak & 116.69 & 158.21 & 0.74 & 1.07 & 1.08 & 1.07 \\
aut & 203.64 & 134.25 & 1.52 & 1.26 & 1.10 & 1.03 \\
isr & 127.57 & 131.37 & 0.97 & 0.79 & 0.86 & 0.98 \\
nor & 387.70 & 124.85 & 3.11 & 1.36 & 1.16 & 1.09 \\
tha & 133.76 & 125.14 & 1.07 & 1.28 & 1.23 & 1.16 \\
grc & 115.30 & 123.25 & 0.94 & 0.98 & 0.93 & 1.01 \\
sgp & 162.48 & 114.10 & 1.42 & 1.13 & 1.15 & 1.18 \\
ukr & 10.10 & 143.64 & 0.07 & 0.63 & 0.68 & 0.66 \\
fin & 229.26 & 109.29 & 2.10 & 1.16 & 1.09 & 1.07 \\
\bottomrule
\end{tabular}
\end{table*}

\clearpage

\begin{table}[htbp]
  \centering
  \caption*{\textbf{Table S2. Top 200 affiliations ranked by the number of papers (published 2015–2023) cited in intergovernmental organization (IGO) policy documents} This table shows institutional contributions to IGO policy-cited research, revealing a concentration in the Anglosphere (USA 59 institutions, GBR 29, AUS 16) and continental Europe (DEU 11, NLD 10, SWE 7, CHE 6). Flagship institutions such as Harvard, Stanford, Oxford, and Wageningen University dominate policy-relevant citations. Outside this core, Asia's presence is led by China's seven entrants—including Tsinghua, Peking, and Fudan—while Japan's University of Tokyo and Korea's Seoul National University appear only once each. Africa is represented chiefly by South Africa (University of Cape Town, Stellenbosch), and Latin America by Brazil's University of São Paulo. Single entries from Singapore, Malaysia, Chile, Mexico, and Ghana highlight how the Global South is represented more by exceptions than by clusters. The geographic footprint demonstrates a densely linked North Atlantic–Australasian corridor supplying most evidence for IGO policy documents, with scattered bridges to the rest of the world.}
  \footnotesize
  \setlength{\tabcolsep}{3pt}

  \begin{minipage}[t]{0.49\linewidth}  
    \vspace{0pt}                      
    \centering
\begin{tabular}{r p{4.5cm} l r}
\toprule
\texttt{\#} & Affiliation & Country & \makecell[c]{\texttt{\#}~Papers} \\
\midrule
1 & \makecell[l]{Wageningen University} & NLD & 1095 \\
2 & \makecell[l]{London School of Hygiene and \\ Tropical Medicine} & GBR & 994 \\
3 & \makecell[l]{University College London} & GBR & 885 \\
4 & \makecell[l]{University of Oxford} & GBR & 878 \\
5 & \makecell[l]{University of Queensland} & AUS & 839 \\
6 & \makecell[l]{World Bank} & USA & 837 \\
7 & \makecell[l]{London School of Economics} & GBR & 817 \\
8 & \makecell[l]{World Health Organization} & CHE & 748 \\
9 & \makecell[l]{University of British Columbia} & CAN & 692 \\
10 & \makecell[l]{Imperial College London} & GBR & 687 \\
11 & \makecell[l]{Stanford University} & USA & 686 \\
12 & \makecell[l]{University of Washington} & USA & 673 \\
13 & \makecell[l]{University of Toronto} & CAN & 633 \\
14 & \makecell[l]{University of Cambridge} & GBR & 619 \\
15 & \makecell[l]{Johns Hopkins University} & USA & 616 \\
16 & \makecell[l]{University of Copenhagen} & DNK & 593 \\
17 & \makecell[l]{Universidade de São Paulo} & BRA & 588 \\
18 & \makecell[l]{University of Leeds} & GBR & 560 \\
19 & \makecell[l]{University of New South Wales} & AUS & 559 \\
20 & \makecell[l]{University of California at \\ Berkeley} & USA & 546 \\
21 & \makecell[l]{University of Michigan} & USA & 543 \\
22 & \makecell[l]{Centers for Disease Control \\ and Prevention} & USA & 540 \\
23 & \makecell[l]{European Commission} & BEL & 538 \\
24 & \makecell[l]{University of Amsterdam} & NLD & 530 \\
25 & \makecell[l]{Harvard School of Public \\ Health} & USA & 524 \\
26 & \makecell[l]{King's College London} & GBR & 522 \\
27 & \makecell[l]{Ghent University} & BEL & 516 \\
28 & \makecell[l]{ETH Zurich} & CHE & 508 \\
29 & \makecell[l]{Columbia University} & USA & 479 \\
30 & \makecell[l]{Aarhus University} & DNK & 477 \\
31 & \makecell[l]{International Monetary Fund} & USA & 473 \\
32 & \makecell[l]{University of Cape Town} & ZAF & 466 \\
33 & \makecell[l]{University of Sydney} & AUS & 464 \\
34 & \makecell[l]{CSIRO} & AUS & 461 \\
35 & \makecell[l]{University of California Davis} & USA & 456 \\
36 & \makecell[l]{McGill University} & CAN & 441 \\
37 & \makecell[l]{University of Melbourne} & AUS & 438 \\
38 & \makecell[l]{Katholieke Universiteit Leuven} & BEL & 426 \\
39 & \makecell[l]{Australian National University} & AUS & 423 \\
40 & \makecell[l]{University of Manchester} & GBR & 423 \\
41 & \makecell[l]{Stockholm University} & SWE & 414 \\
42 & \makecell[l]{Michigan State University} & USA & 411 \\
43 & \makecell[l]{Cornell University} & USA & 410 \\
44 & \makecell[l]{Yale University} & USA & 408 \\
45 & \makecell[l]{University of Minnesota} & USA & 407 \\
46 & \makecell[l]{University of Maryland} & USA & 401 \\
47 & \makecell[l]{Harvard University} & USA & 397 \\
48 & \makecell[l]{University of Helsinki} & FIN & 394 \\
49 & \makecell[l]{National University of \\ Singapore} & SGP & 389 \\
50 & \makecell[l]{University of Edinburgh} & GBR & 386 \\
\bottomrule
\end{tabular}

  \end{minipage}
  \hfill
 \begin{minipage}[t]{0.49\linewidth} 
    \vspace{0pt}
    \centering
\begin{tabular}{r p{4.5cm} l r}
\toprule
\texttt{\#} & Affiliation & Country & \makecell[c]{\texttt{\#}~Papers} \\
\midrule
51 & \makecell[l]{Massachusetts Institute of \\ Technology} & USA & 382 \\
52 & \makecell[l]{Tsinghua University} & CHN & 381 \\
53 & \makecell[l]{Monash University} & AUS & 378 \\
54 & \makecell[l]{University of Exeter} & GBR & 374 \\
55 & \makecell[l]{University of Oslo} & NOR & 366 \\
56 & \makecell[l]{University of Southampton} & GBR & 364 \\
57 & \makecell[l]{University of Wisconsin} & USA & 364 \\
58 & \makecell[l]{Griffith University} & AUS & 357 \\
59 & \makecell[l]{International Food Policy \\ Research Institute} & USA & 356 \\
60 & \makecell[l]{University of Gothenburg} & SWE & 352 \\
61 & \makecell[l]{University of Oxford} & GBR & 348 \\
62 & \makecell[l]{University of Illinois} & USA & 344 \\
63 & \makecell[l]{Karolinska Institutet} & SWE & 341 \\
64 & \makecell[l]{University of Sussex} & GBR & 340 \\
65 & \makecell[l]{Peking University} & CHN & 338 \\
66 & \makecell[l]{University of Groningen} & NLD & 332 \\
67 & \makecell[l]{Technical University of \\ Denmark} & DNK & 330 \\
68 & \makecell[l]{Duke University} & USA & 330 \\
69 & \makecell[l]{University of York} & GBR & 328 \\
70 & \makecell[l]{University of North Carolina} & USA & 328 \\
71 & \makecell[l]{Utrecht University} & NLD & 326 \\
72 & \makecell[l]{University of Alberta} & CAN & 321 \\
73 & \makecell[l]{Pennsylvania State University} & USA & 320 \\
74 & \makecell[l]{University of California Los \\ Angeles} & USA & 318 \\
75 & \makecell[l]{Princeton University} & USA & 318 \\
76 & \makecell[l]{James Cook University} & AUS & 317 \\
77 & \makecell[l]{University of Birmingham} & GBR & 316 \\
78 & \makecell[l]{Norwegian University of \\ Science and Technology} & NOR & 316 \\
79 & \makecell[l]{University of Bern} & CHE & 309 \\
80 & \makecell[l]{Lund University} & SWE & 307 \\
81 & \makecell[l]{Vrije Universiteit Amsterdam} & NLD & 307 \\
82 & \makecell[l]{University of Ottawa} & CAN & 305 \\
83 & \makecell[l]{Ohio State University} & USA & 304 \\
84 & \makecell[l]{University of Zurich} & CHE & 304 \\
85 & \makecell[l]{University of Bristol} & GBR & 299 \\
86 & \makecell[l]{University of Göttingen} & DEU & 289 \\
87 & \makecell[l]{Swedish University of \\ Agricultural Sciences} & SWE & 286 \\
88 & \makecell[l]{University of Antwerp} & BEL & 285 \\
89 & \makecell[l]{Oregon State University} & USA & 283 \\
90 & \makecell[l]{University of Reading} & GBR & 279 \\
91 & \makecell[l]{Beijing Normal University} & CHN & 279 \\
92 & \makecell[l]{University of Colorado} & USA & 274 \\
93 & \makecell[l]{University of Waterloo} & CAN & 273 \\
94 & \makecell[l]{Deakin University} & AUS & 272 \\
95 & \makecell[l]{Uppsala University} & SWE & 271 \\
96 & \makecell[l]{Chinese Academy of Sciences} & CHN & 268 \\
97 & \makecell[l]{University of Auckland} & NZL & 266 \\
98 & \makecell[l]{University of Texas at Austin} & USA & 266 \\
99 & \makecell[l]{University of Pennsylvania} & USA & 264 \\
100 & \makecell[l]{University of Nottingham} & GBR & 263 \\
\bottomrule
\end{tabular}

  \end{minipage}

\end{table}

\begin{table}[htbp]
  \centering
  \captionsetup{justification=centering}
  \footnotesize
  \setlength{\tabcolsep}{3pt}

  \begin{minipage}[t]{0.49\linewidth}   
    \vspace{0pt}                        
    \centering
\begin{tabular}{r p{4.5cm} l r}
\toprule
\texttt{\#} & Affiliation & Country & \makecell[c]{\texttt{\#}~Papers} \\
\midrule
101 & \makecell[l]{New York University} & USA & 263 \\
102 & \makecell[l]{University of Hong Kong} & HKG & 257 \\
103 & \makecell[l]{University of Chicago} & USA & 256 \\
104 & \makecell[l]{American University of Beirut} & USA & 254 \\
105 & \makecell[l]{Arizona State University} & USA & 254 \\
106 & \makecell[l]{Ludwig-Maximilians-Universität \\ München} & DEU & 251 \\
107 & \makecell[l]{University of Florida} & USA & 247 \\
108 & \makecell[l]{University of Liverpool} & GBR & 247 \\
109 & \makecell[l]{University of Oxford} & GBR & 246 \\
110 & \makecell[l]{Harvard Medical School} & USA & 246 \\
111 & \makecell[l]{University of California San \\ Francisco} & USA & 246 \\
112 & \makecell[l]{University of Western \\ Australia} & AUS & 243 \\
113 & \makecell[l]{Leiden University} & NLD & 243 \\
114 & \makecell[l]{Texas A and M University} & USA & 243 \\
115 & \makecell[l]{University of California at \\ Santa Barbara} & USA & 243 \\
116 & \makecell[l]{University of Southern \\ California} & USA & 241 \\
117 & \makecell[l]{Universitat de Barcelona} & ESP & 239 \\
118 & \makecell[l]{University of KwaZulu-Natal} & ZAF & 239 \\
119 & \makecell[l]{University of Pretoria} & ZAF & 237 \\
120 & \makecell[l]{University of Tokyo} & JPN & 235 \\
121 & \makecell[l]{Colorado State University} & USA & 234 \\
122 & \makecell[l]{University of Adelaide} & AUS & 232 \\
123 & \makecell[l]{University of Arizona} & USA & 230 \\
124 & \makecell[l]{University of California \\ Irvine} & USA & 228 \\
125 & \makecell[l]{University of Sheffield} & GBR & 222 \\
126 & \makecell[l]{University of Bergen} & NOR & 219 \\
127 & \makecell[l]{Seoul National University} & KOR & 219 \\
128 & \makecell[l]{Umeå University} & SWE & 217 \\
129 & \makecell[l]{McMaster University} & CAN & 216 \\
130 & \makecell[l]{U.S. Geological Survey} & USA & 215 \\
131 & \makecell[l]{Queensland University of \\ Technology} & AUS & 213 \\
132 & \makecell[l]{University of East Anglia} & GBR & 213 \\
133 & \makecell[l]{Food and Agriculture \\ Organization of the United} & ITA & 211 \\
134 & \makecell[l]{University of Hamburg} & DEU & 211 \\
135 & \makecell[l]{Technische Universität München} & DEU & 210 \\
136 & \makecell[l]{Macquarie University} & AUS & 210 \\
137 & \makecell[l]{Universidade Federal do Rio de \\ Janeiro} & BRA & 208 \\
138 & \makecell[l]{Boston University} & USA & 206 \\
139 & \makecell[l]{University of Bologna} & ITA & 206 \\
140 & \makecell[l]{Institute of Geographic \\ Sciences and Natural Resources} & CHN & 205 \\
141 & \makecell[l]{University of Otago} & NZL & 202 \\
142 & \makecell[l]{Curtin University} & AUS & 202 \\
143 & \makecell[l]{University of Warwick} & GBR & 201 \\
144 & \makecell[l]{University of Lausanne} & CHE & 201 \\
145 & \makecell[l]{Universidad de Chile} & CHL & 201 \\
146 & \makecell[l]{University of Ghana} & GHA & 199 \\
147 & \makecell[l]{Stellenbosch University} & ZAF & 198 \\
148 & \makecell[l]{CNRS} & FRA & 198 \\
149 & \makecell[l]{University of the \\ Witwatersrand} & ZAF & 197 \\
150 & \makecell[l]{Simon Fraser University} & CAN & 196 \\
\bottomrule
\end{tabular}

  \end{minipage}
  \hfill 
  \begin{minipage}[t]{0.49\linewidth}
    \vspace{0pt}
    \centering
\begin{tabular}{r p{4.5cm} l r}
\toprule
\texttt{\#} & Affiliation & Country & \makecell[c]{\texttt{\#}~Papers} \\
\midrule
151 & \makecell[l]{Purdue University} & USA & 196 \\
152 & \makecell[l]{Dalhousie University} & CAN & 194 \\
153 & \makecell[l]{University of Malaya} & MYS & 194 \\
154 & \makecell[l]{Alfred Wegener Institute for \\ Polar and Marine Research} & DEU & 194 \\
155 & \makecell[l]{International Institute for \\ Applied Systems Analysis} & AUT & 193 \\
156 & \makecell[l]{Maastricht University} & NLD & 191 \\
157 & \makecell[l]{University of Virginia} & USA & 191 \\
158 & \makecell[l]{University of Technology \\ Sydney} & AUS & 191 \\
159 & \makecell[l]{Sun Yat-sen University} & CHN & 191 \\
160 & \makecell[l]{University of Geneva} & CHE & 190 \\
161 & \makecell[l]{Norwegian University of Life \\ Sciences} & NOR & 189 \\
162 & \makecell[l]{University College Dublin} & IRL & 189 \\
163 & \makecell[l]{Washington University} & USA & 188 \\
164 & \makecell[l]{Tufts University} & USA & 188 \\
165 & \makecell[l]{University of Padova} & ITA & 188 \\
166 & \makecell[l]{Universidade Federal de Minas \\ Gerais} & BRA & 187 \\
167 & \makecell[l]{Pontificia Universidad \\ Católica de Chile} & CHL & 186 \\
168 & \makecell[l]{Fudan University} & CHN & 185 \\
169 & \makecell[l]{Universidad Nacional Autónoma \\ de México} & MEX & 185 \\
170 & \makecell[l]{Northwestern University} & USA & 184 \\
171 & \makecell[l]{Columbia University} & USA & 183 \\
172 & \makecell[l]{Potsdam Institute for Climate \\ Impact Research} & DEU & 183 \\
173 & \makecell[l]{Brown University} & USA & 183 \\
174 & \makecell[l]{Utrecht University} & NLD & 183 \\
175 & \makecell[l]{Erasmus University Medical \\ Center} & NLD & 183 \\
176 & \makecell[l]{Indiana University} & USA & 182 \\
177 & \makecell[l]{Universitat Autònoma de \\ Barcelona} & ESP & 182 \\
178 & \makecell[l]{Sapienza University of Rome} & ITA & 182 \\
179 & \makecell[l]{University of Western Ontario} & CAN & 182 \\
180 & \makecell[l]{University of Glasgow} & GBR & 182 \\
181 & \makecell[l]{University of Heidelberg} & DEU & 180 \\
182 & \makecell[l]{Université Catholique de \\ Louvain} & BEL & 180 \\
183 & \makecell[l]{Freie Universität Berlin} & DEU & 179 \\
184 & \makecell[l]{Karlsruhe Institute of \\ Technology} & DEU & 178 \\
185 & \makecell[l]{Inter-American Development \\ Bank} & USA & 178 \\
186 & \makecell[l]{University of California San \\ Diego} & USA & 178 \\
187 & \makecell[l]{Public Health England} & GBR & 176 \\
188 & \makecell[l]{OECD} & FRA & 175 \\
189 & \makecell[l]{Cardiff University} & GBR & 174 \\
190 & \makecell[l]{Goethe University Frankfurt} & DEU & 173 \\
191 & \makecell[l]{Johns Hopkins University} & USA & 173 \\
192 & \makecell[l]{Met Office} & GBR & 172 \\
193 & \makecell[l]{Instituto Nacional de Salud \\ Pública} & MEX & 170 \\
194 & \makecell[l]{University of Milan} & ITA & 169 \\
195 & \makecell[l]{University of Twente} & NLD & 169 \\
196 & \makecell[l]{Johns Hopkins University \\ School of Medicine} & USA & 169 \\
197 & \makecell[l]{Newcastle University} & GBR & 168 \\
198 & \makecell[l]{U.S. Environmental Protection \\ Agency} & USA & 168 \\
199 & \makecell[l]{Université de Montpellier} & FRA & 167 \\
200 & \makecell[l]{Technische Universität Berlin} & DEU & 166 \\
\bottomrule
\end{tabular}

  \end{minipage}

\end{table}

\clearpage

\begin{figure}[!h]
\centering
\includegraphics[width=0.5\textwidth]{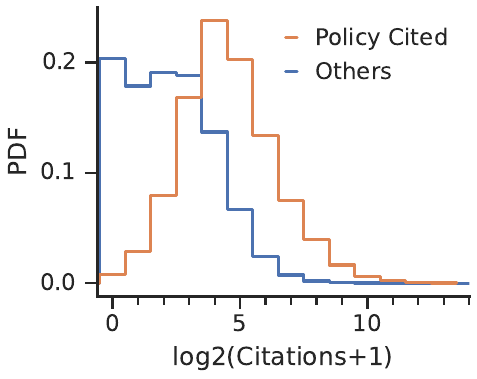}
\caption*{\textbf{Figure S1. Academic impact comparison between policy-cited and non-policy-cited papers published in 2020} Probability density functions of $\log_2(\text{citations} + 1)$
 for papers published in 2020, based on academic citations accumulated through the end of 2023. Orange curve represents policy-cited papers; blue curve represents non-policy-cited papers. Policy-cited papers show systematically higher academic citations, with their distribution peak shifted rightward on the log-transformed citation scale. This pattern confirms that policy influence correlates with academic impact, suggesting that policymakers preferentially cite scientifically influential research. The rightward shift of the orange curve indicates that papers cited in policy documents tend to receive more academic citations than those not cited in policy, demonstrating a positive relationship between policy relevance and scientific influence within the academic community.}
\end{figure}

\clearpage

\begin{table*}[ht]
\footnotesize
\centering
\caption*{\textbf{Table S3. Summary of 23 clusters of policy-cited papers identified through citation network analysis.} Clusters were identified using Leiden Algorithm on the citation network of policy-cited papers. High-frequency TF-IDF terms were extracted and GPT-4 generated descriptive labels. The table reports: cluster category, name, number of papers, defining keywords, leading countries, top institutions, and most influential authors by policy-citation metrics. Clusters span three domains: Environmental (conservation, energy, marine, climate, pollution), Health (healthcare, COVID-19, infectious diseases, public health), and Social Sciences (economy, AI, urban planning). Cluster sizes range from ~1,300 to 16,000+ papers, reflecting differential policy attention across topics.}
\begin{tabular}{|p{1.8cm}|p{2.5cm}|p{1.1cm}|p{3.2cm}|p{2.1cm}|p{4.7cm}|}
\hline
\textbf{Category} & \textbf{Cluster Name} & \textbf{\texttt{\#}Papers} & \textbf{Top TF-IDF Words} & \textbf{Top Countries} & \textbf{Top Institutions} \\
\hline
\hline

\textbf{Environment} & \textbf{Environmental Conservation} & 16358 & land, water, forest, soil, food, species, biodiversity, conservation, production, ecosystem, agricultural, management, use, environmental, climate & usa, gbr, chn, deu, aus, nld, bra, can, fra, ita & Wageningen University, University of Oxford, University of Queensland, European Commission, Swedish University of Agricultural Sciences, CSIRO, University of Copenhagen, Chinese Academy of Sciences, ETH Zurich, University of Maryland \\
        \multicolumn{6}{|p{17.7cm}|}{\textbf{Top Authors:} Barbier Edward B.: Colorado State University (USA), Hoekstra Arjen Y.: University of Twente (NLD), Lambin Eric F.: Stanford University (USA), Clapp Jennifer: University of Waterloo (CAN), Verburg Peter H.: Vrije Universiteit Amsterdam (NLD), Fearnside Philip M.: Instituto Nacional de Pesquisas da Amazônia (BRA), Watson James E. M.: University of Queensland (AUS), Herrero Mario: CSIRO (AUS), Panagos Panos: European Commission (BEL), Baird Ian G.: University of Wisconsin (USA) } \\
        \hline
\textbf{Environment} & \textbf{Energy Policy} & 13250 & energy, carbon, emissions, climate, electricity, policy, renewable, power, co, environmental, change, economic, global, soil, countries & usa, gbr, chn, deu, aus, nld, swe, fra, esp, can & ETH Zurich, Tsinghua University, European Commission, University College London, University of Oxford, International Institute for Applied Systems Analysis, Utrecht University, Imperial College London, University of Leeds, University of Cambridge \\
        \multicolumn{6}{|p{17.7cm}|}{\textbf{Top Authors:} Sovacool Benjamin K.: Aarhus University (DNK), van Vuuren Detlef P.: PBL Netherlands Environmental Assessment Agency (NLD), Breyer Christian: Lappeenranta University of Technology (FIN), van der Ploeg Frederick: Vrije Universiteit Amsterdam (NLD), Jakob Michael: Mercator Research Institute on Global Commons and Climate Change (DEU), Fujimori Shinichiro: National Institute for Environmental Studies (JPN), Urpelainen Johannes: Johns Hopkins University (USA), Schmidt Tobias S.: ETH Zurich (CHE), Hertwich Edgar G.: Yale University (USA), Creutzig Felix: Technische Universität Berlin (DEU) } \\
        \hline
\textbf{Environment} & \textbf{Marine Conservation} & 12621 & species, marine, fisheries, sea, ocean, fish, climate, coastal, fishing, change, management, whales, ecosystem, coral, conservation & usa, aus, gbr, can, chn, deu, fra, esp, nor, ita & University of British Columbia, University of Washington, James Cook University, University of Tasmania, CSIRO, University of Queensland, NOAA, Institute of Marine Research, Wageningen University, Oregon State University \\
        \multicolumn{6}{|p{17.7cm}|}{\textbf{Top Authors:} Sumaila U. Rashid: University of British Columbia (CAN), Cheung William W.L.: University of British Columbia (CAN), Duarte Carlos M.: King Abdullah University of Science and Technology (KAUST) (SAU), Pauly Daniel: University of British Columbia (CAN), Friedlaender Ari S.: University of California Santa Cruz (USA), Friess Daniel A.: National University of Singapore (SGP), Punt André E.: University of Washington (USA), Lovelock Catherine E.: University of Queensland (AUS), Levin Lisa A.: University of California San Diego (USA), Mercier Annie: Memorial University of Newfoundland (CAN) } \\
        \hline
\textbf{Environment} & \textbf{Climate Modeling} & 12485 & climate, precipitation, ice, model, temperature, change, global, changes, warming, water, sea, surface, drought, models, atmospheric & usa, gbr, chn, deu, fra, aus, can, che, nld, esp & Met Office, National Center for Atmospheric Research, Columbia University, University of Reading, University of Colorado, Institute for Atmospheric and Climate Science, University of Oxford, NOAA, Chinese Academy of Sciences, University of Leeds \\
        \multicolumn{6}{|p{17.7cm}|}{\textbf{Top Authors:} Seneviratne Sonia I.: Institute for Atmospheric and Climate Science (CHE), Almazroui Mansour: King Abdulaziz University (SAU), Ciais Philippe: Université Paris-Saclay (FRA), Wehner Michael F.: Lawrence Berkeley National Laboratory (USA), Xie Shang-Ping: University of California San Diego (USA), Fischer Erich M.: Institute for Atmospheric and Climate Science (CHE), Aerts Jeroen C. J. H.: Vrije Universiteit Amsterdam (NLD), Giorgi Filippo: International Centre for Theoretical Physics (ITA), Wada Yoshihide: International Institute for Applied Systems Analysis (AUT), Otto Friederike E. L.: University of Oxford (GBR) } \\
        \hline
\textbf{Environment} & \textbf{Urban Mobility} & 5096 & air, pm, pollution, transport, travel, exposure, urban, health, traffic, transit, mobility, cities, use, car, data & usa, gbr, chn, aus, can, nld, deu, swe, esp, fra & University College London, University of California at Berkeley, Delft University of Technology, University of Leeds, Peking University, Tsinghua University, Columbia University, University of Queensland, University of Sydney, Universidad de Los Andes \\
        \multicolumn{6}{|p{17.7cm}|}{\textbf{Top Authors:} Guzman Luis A.: Universidad de Los Andes (COL), Oviedo Daniel: University College London (GBR), Börjesson Maria: Royal Institute of Technology (SWE), Nieuwenhuijsen Mark J.: Universitat de Barcelona (ESP), Aldred Rachel: University of Westminster (GBR), van Wee Bert: Delft University of Technology (NLD), Mulley Corinne: University of Sydney (AUS), Hensher David A.: University of Sydney (AUS), Bertolini Luca: University of Amsterdam (NLD), Brauer Michael: University of British Columbia (CAN) } \\
        \hline

\end{tabular}
\label{tab:environmental_summary}
\end{table*}

\begin{table*}[ht]
\footnotesize
\centering
\begin{tabular}{|p{1.8cm}|p{2.5cm}|p{1.1cm}|p{3.2cm}|p{2.1cm}|p{4.7cm}|}
\hline
\textbf{Category} & \textbf{Cluster Name} & \textbf{\texttt{\#}Papers} & \textbf{Top TF-IDF Words} & \textbf{Top Countries} & \textbf{Top Institutions} \\
\hline
\hline

\textbf{Environment} & \textbf{Plastic Pollution} & 5075 & plastic, microplastics, marine, microplastic, plastics, waste, exposure, litter, concentrations, debris, chemicals, mps, particles, pollution, environmental & usa, chn, gbr, deu, can, ita, nld, esp, aus, fra & U.S. Environmental Protection Agency, Wageningen University, Chinese Academy of Sciences, Technical University of Denmark, University of Plymouth, East China Normal University, Stockholm University, Environment Canada, University of Hong Kong, Vrije Universiteit Amsterdam \\
        \multicolumn{6}{|p{17.7cm}|}{\textbf{Top Authors:} Walker Tony R.: Dalhousie University (CAN), Thompson Richard C.: University of Plymouth (GBR), Turner Andrew: University of Plymouth (GBR), Hardesty Britta Denise: CSIRO (AUS), Koelmans Albert A.: Wageningen University (NLD), Wilcox Chris: CSIRO (AUS), Weber Roland: POPs Environmental Consulting (DEU), Ryan Peter G.: University of Cape Town (ZAF), Kannan Kurunthachalam: University at Albany (USA), Fiedler Heidelore: Örebro University (SWE) } \\
        \hline
\textbf{Environment} & \textbf{Circular Economy} & 2842 & waste, food, circular, recycling, environmental, economy, management, ce, blockchain, supply, sustainable, chain, sustainability, energy, study & usa, chn, gbr, ita, deu, nld, esp, bra, swe, aus & Wageningen University, Shantou University, Tsinghua University, University of Cambridge, Utrecht University, Lund University, Technical University of Denmark, University of Manchester, Delft University of Technology, European Commission \\
        \multicolumn{6}{|p{17.7cm}|}{\textbf{Top Authors:} Pearce Joshua M.: Michigan Technological University (USA), Huo Xia: Jinan University (CHN), Xu Xijin: Shantou University (CHN), Kshetri Nir: University of North Carolina (USA), Li Jinhui: Tsinghua University (CHN), Bocken Nancy: Delft University of Technology (NLD), Gutberlet Jutta: University of Victoria (CAN), Sumida Huaman Elizabeth: University of Minnesota (USA), Tukker Arnold: Leiden University (NLD), Milios Leonidas: Lund University (SWE) } \\
        \hline
\textbf{Environment} & \textbf{Environmental Toxicology} & 1333 & hg, mercury, mining, water, mehg, concentrations, desalination, asm, soil, gold, artisanal, asgm, metals, environmental, fish & usa, chn, can, aus, gbr, swe, bra, esp, deu, fra & University of Michigan, Tsinghua University, University of Queensland, McGill University, U.S. Geological Survey, University of British Columbia, Duke University, University of Melbourne, University of Connecticut, Institute of Geochemistry \\
        \multicolumn{6}{|p{17.7cm}|}{\textbf{Top Authors:} Hilson Gavin: University of Surrey (GBR), Lahiri-Dutt Kuntala: Australian National University (AUS), Verbrugge Boris: Radboud University (NLD), Vanclay Frank: University of Groningen (NLD), Mason Robert P.: University of Connecticut (USA), Gude Veera Gnaneswar: Mississippi State University (USA), Basu Niladri: McGill University (CAN), Spiegel Samuel J.: University of Edinburgh (GBR), Zetterberg Pär: Uppsala University (SWE), Larsen Rasmus Kløcker: Stockholm Environment Institute (SWE) } \\
        \hline
\textbf{Health Care} & \textbf{Healthcare} & 14741 & health, care, cancer, patients, studies, quality, healthcare, countries, services, study, women, data, evidence, methods, research & usa, gbr, aus, can, chn, nld, che, deu, zaf, ind & World Health Organization, London School of Hygiene and Tropical Medicine, Johns Hopkins University, University College London, University of Toronto, University of Oxford, Harvard School of Public Health, University of Sydney, King's College London, McMaster University \\
        \multicolumn{6}{|p{17.7cm}|}{\textbf{Top Authors:} Ridde Valéry: Université de Montréal (CAN), Fancourt Daisy: University College London (GBR), Kuper Hannah: London School of Hygiene and Tropical Medicine (GBR), McKee Martin: London School of Hygiene and Tropical Medicine (GBR), George Asha: University of the Western Cape (ZAF), Barasa Edwine: KEMRI-Wellcome Trust Research Programme (KEN), Gilson Lucy: University of Cape Town (ZAF), Patel Vikram: Harvard Medical School (USA), Witter Sophie: Queen Margaret University (GBR), Steptoe Andrew: University College London (GBR) } \\
        \hline
\textbf{Health Care} & \textbf{Covid19 Pandemic} & 12190 & covid, 19, pandemic, cov, sars, patients, health, disease, coronavirus, 2020, infection, 95, care, ci, respiratory & usa, gbr, chn, ita, can, deu, aus, ind, esp, fra & University of Oxford, Centers for Disease Control and Prevention, London School of Hygiene and Tropical Medicine, University College London, World Health Organization, University of Toronto, University of Hong Kong, Imperial College London, Columbia University, Johns Hopkins University \\
        \multicolumn{6}{|p{17.7cm}|}{\textbf{Top Authors:} Mahase Elisabeth: The BMJ (GBR), Iacobucci Gareth: The BMJ (GBR), Dyer Owen: NO DATA (NO DATA), McKee Martin: London School of Hygiene and Tropical Medicine (GBR), Ludvigsson Jonas F.: Örebro University (SWE), Lippi Giuseppe: University of Verona (ITA), Krammer Florian: Icahn School of Medicine at Mount Sinai (USA), Al-Tawfiq Jaffar A.: Johns Hopkins Aramco Healthcare (SAU), Lipsitch Marc: Harvard School of Public Health (USA), Holt-Lunstad Julianne: Brigham Young University (USA) } \\
        \hline

\end{tabular}
\label{tab:environmental_summary}
\end{table*}

\begin{table*}[ht]
\footnotesize
\centering
\begin{tabular}{|p{1.8cm}|p{2.5cm}|p{1.1cm}|p{3.2cm}|p{2.1cm}|p{4.7cm}|}
\hline
\textbf{Category} & \textbf{Cluster Name} & \textbf{\texttt{\#}Papers} & \textbf{Top TF-IDF Words} & \textbf{Top Countries} & \textbf{Top Institutions} \\
\hline
\hline

\textbf{Health Care} & \textbf{Infectious Diseases (excl. Covid19)} & 7371 & hiv, tb, health, treatment, patients, care, tuberculosis, hcv, 95, ci, among, testing, women, drug, risk & usa, gbr, zaf, aus, can, chn, deu, che, nld, swe & World Health Organization, Centers for Disease Control and Prevention, London School of Hygiene and Tropical Medicine, University of Cape Town, Johns Hopkins University, University of the Witwatersrand, University of California San Francisco, University of Washington, McGill University, University of KwaZulu-Natal \\
        \multicolumn{6}{|p{17.7cm}|}{\textbf{Top Authors:} Pai Madhukar: McGill University (CAN), Abu-Raddad Laith J.: Cornell University (QAT), Migliori Giovanni Battista: IRCCS (ITA), Ford Nathan: World Health Organization (CHE), Triandafyllidou Anna: Ryerson University (CAN), Mayer Kenneth H.: Harvard Medical School (USA), Norredam Marie: University of Copenhagen (DNK), Harries Anthony D.: London School of Hygiene and Tropical Medicine (GBR), Lönnroth Knut: Karolinska Institutet (SWE), Dowdy David W.: Johns Hopkins University (USA) } \\
        \hline
\textbf{Health Care} & \textbf{Antibiotic Resistance} & 5644 & antibiotic, resistance, antibiotics, antimicrobial, isolates, amr, health, bacteria, resistant, food, fish, gut, use, coli, water & usa, chn, gbr, aus, can, fra, nld, deu, ita, esp & Centers for Disease Control and Prevention, World Health Organization, University of Oxford, Food and Drug Administration, London School of Hygiene and Tropical Medicine, Mahidol University, China Agricultural University, Wageningen University, Kasetsart University, Ghent University \\
        \multicolumn{6}{|p{17.7cm}|}{\textbf{Top Authors:} Surachetpong Win: Kasetsart University (THA), Dong Ha Thanh: Suan Sunandha Rajabhat University (THA), Laxminarayan Ramanan: Princeton University (USA), Senapin Saengchan: Mahidol University (THA), Allende Ana: CSIC (ESP), Crump John A.: University of Otago (NZL), Outterson Kevin: Boston University (USA), Larsson D.G. Joakim: University of Gothenburg (SWE), Fan Xuetong: Eastern Regional Research Center (USA), Rushton Jonathan: University of Liverpool (GBR) } \\
        \hline
\textbf{Health Care} & \textbf{Public Health} & 5543 & alcohol, food, obesity, health, consumption, intake, diabetes, risk, foods, 95, tobacco, children, ci, dietary, studies & usa, gbr, aus, can, chn, nld, deu, bra, fra, esp & Deakin University, Instituto Nacional de Salud Pública, University of Sydney, University of North Carolina, University of Toronto, London School of Hygiene and Tropical Medicine, Harvard School of Public Health, University of Cambridge, University of Oxford, Johns Hopkins University \\
        \multicolumn{6}{|p{17.7cm}|}{\textbf{Top Authors:} Rehm Jürgen: University of Toronto (CAN), Popkin Barry M.: University of North Carolina (USA), Sacks Gary: Deakin University (AUS), Mozaffarian Dariush: Tufts University (USA), Swinburn Boyd: University of Auckland (NZL), Vandevijvere Stefanie: University of Auckland (NZL), Chaloupka Frank J.: University of Illinois (USA), Gilmore Anna B.: University of Bath (GBR), Rivera Juan A.: Instituto Nacional de Salud Pública (MEX), Ng Shu Wen: University of North Carolina (USA) } \\
        \hline
\textbf{Health Care} & \textbf{Drug Forensics} & 4841 & drug, ms, samples, forensic, detection, drugs, cannabis, method, use, mass, cocaine, spectrometry, analysis, ng, compounds & usa, chn, gbr, aus, deu, ind, bra, ita, esp, irn & University of Lausanne, University of Technology Sydney, National Institute of Standards and Technology, Netherlands Forensic Institute, University of Queensland, University of New South Wales, King's College London, Universiti Sains Malaysia, University of Copenhagen, University College London \\
        \multicolumn{6}{|p{17.7cm}|}{\textbf{Top Authors:} García-Ruiz Carmen: Universidad de Alcalá (ESP), Nagabhushana Hanumanthappa: Tumkur University (IND), Roux Claude: University of Technology Sydney (AUS), Meyer Markus R.: Saarland University (DEU), Palamar Joseph J.: New York University (USA), Huestis Marilyn A.: National Institute on Drug Abuse (USA), Maurer Hans H.: Saarland University (DEU), Darshan Giriyapura Prabhukumar: Acharya Institute of Graduate Studies (IND), Beck Olof: Karolinska Institutet (SWE), Brandt Simon D.: Liverpool John Moores University (GBR) } \\
        \hline
\textbf{Health Care} & \textbf{Vector-Borne Diseases} & 4446 & malaria, dengue, vector, control, transmission, zikv, infection, disease, health, zika, virus, gene, genome, resistance, mosquito & usa, gbr, chn, bra, che, aus, ind, fra, deu, nld & London School of Hygiene and Tropical Medicine, World Health Organization, Centers for Disease Control and Prevention, University of Oxford, Fundação Oswaldo Cruz, Imperial College London, Liverpool School of Tropical Medicine, University of Edinburgh, University of California San Francisco, Johns Hopkins University \\
        \multicolumn{6}{|p{17.7cm}|}{\textbf{Top Authors:} Solomon Anthony W.: London School of Hygiene and Tropical Medicine (GBR), Gao Caixia: Chinese Academy of Sciences (CHN), Lindsay Steve W.: Durham University (GBR), Chimbari Moses J.: University of KwaZulu-Natal (ZAF), Wesseler Justus: Wageningen University (NLD), Torr Stephen J.: Liverpool School of Tropical Medicine (GBR), Kuzma Jennifer: North Carolina State University (USA), Rodriguez-Morales Alfonso J.: Universidad Tecnológica de Pereira (COL), De Vlas Sake J.: Erasmus University Medical Center (NLD), Hargrove John W.: Stellenbosch University (ZAF) } \\
        \hline

\end{tabular}
\label{tab:environmental_summary}
\end{table*}

\begin{table*}[ht]
\footnotesize
\centering
\begin{tabular}{|p{1.8cm}|p{2.5cm}|p{1.1cm}|p{3.2cm}|p{2.1cm}|p{4.7cm}|}
\hline
\textbf{Category} & \textbf{Cluster Name} & \textbf{\texttt{\#}Papers} & \textbf{Top TF-IDF Words} & \textbf{Top Countries} & \textbf{Top Institutions} \\
\hline
\hline

\textbf{Health Care} & \textbf{Animal Diseases} & 1535 & virus, fmd, rabies, disease, fmdv, asf, vaccine, viruses, wild, vaccination, influenza, animals, pigs, animal, cattle & usa, chn, gbr, deu, aus, ind, ita, kor, nld, fra & Friedrich-Loeffler-Institut, Institute for Animal Health, Centers for Disease Control and Prevention, Animal and Plant Quarantine Agency, Chinese Academy of Agricultural Sciences, CSIRO, Wageningen University, Plum Island Animal Disease Center, University of London, University of Pretoria \\
        \multicolumn{6}{|p{17.7cm}|}{\textbf{Top Authors:} Beer Martin: Friedrich-Loeffler-Institut (DEU), Blome Sandra: Friedrich-Loeffler-Institut (DEU), Rupprecht Charles E.: LYSSA LLC (USA), Pfeiffer Dirk U.: University of London (GBR), Wallace Ryan M.: Centers for Disease Control and Prevention (USA), King Donald P.: Institute for Animal Health (GBR), Park Jong-Hyeon: Animal and Plant Quarantine Agency (KOR), Ståhl Karl: National Veterinary Institute (SWE), Beltrán-Alcrudo Daniel: Regional Office for Europe and Central Asia (HUN), Depner Klaus: Friedrich-Loeffler-Institut (DEU) } \\
        \hline
\textbf{Health Care} & \textbf{Tobacco Control} & 1236 & cigarette, tobacco, cigarettes, smoking, nicotine, smokers, use, products, waterpipe, smoke, users, exposure, menthol, among, cessation & usa, gbr, can, chn, che, irn, aus, deu, kor, jpn & University of Minnesota, University of California San Francisco, University of Southern California, Virginia Commonwealth University, British American Tobacco, Johns Hopkins University, American University of Beirut, Center for Tobacco Products, Philip Morris Products S.A., University of Waterloo \\
        \multicolumn{6}{|p{17.7cm}|}{\textbf{Top Authors:} Glantz Stanton A.: University of California San Francisco (USA), Hecht Stephen S.: University of Minnesota (USA), Benowitz Neal L.: University of California San Francisco (USA), Eissenberg Thomas: Virginia Commonwealth University (USA), Jawad Mohammed: Imperial College London (GBR), Goniewicz Maciej L.: Roswell Park Cancer Institute (USA), Hatsukami Dorothy K.: University of Minnesota (USA), Leventhal Adam M.: University of Southern California (USA), Delnevo Cristine D.: Rutgers University (USA), Shihadeh Alan: American University of Beirut (USA) } \\
        \hline
\textbf{Social Science} & \textbf{Economy and Finance} & 21218 & firms, countries, financial, trade, growth, find, economic, policy, market, firm, innovation, data, effects, model, capital & usa, gbr, deu, chn, ita, fra, nld, esp, can, che & International Monetary Fund, World Bank, London School of Economics, National Bureau of Economic Research, Harvard University, University of Oxford, University of Chicago, Stanford University, Federal Reserve Board, Columbia University \\
        \multicolumn{6}{|p{17.7cm}|}{\textbf{Top Authors:} Rodríguez-Pose Andrés: London School of Economics (GBR), Jalles João Tovar: Universidade Nova de Lisboa (PRT), Yoshino Naoyuki: Keio University (JPN), Hoekman Bernard: European University Institute (ITA), Acemoglu Daron: Massachusetts Institute of Technology (USA), Furceri Davide: International Monetary Fund (USA), Taghizadeh-Hesary Farhad: Tokai University (JPN), Binder Carola: Haverford College (USA), Boschma Ron: Utrecht University (NLD), de Haan Jakob: University of Groningen (NLD) } \\
        \hline
\textbf{Social Science} & \textbf{Education System} & 12696 & school, students, children, education, teachers, learning, schools, research, educational, teacher, social, study, gender, student, data & usa, gbr, aus, deu, can, chn, nld, esp, swe, nor & Harvard University, Stanford University, Columbia University, University of London, University of Oxford, University of Cambridge, University of Melbourne, University of Toronto, University of Hong Kong, University of Pennsylvania \\
        \multicolumn{6}{|p{17.7cm}|}{\textbf{Top Authors:} Wodon Quentin: World Bank (USA), Williamson Ben: University of Stirling (GBR), Wolf Sharon: University of Pennsylvania (USA), Przybylski Andrew K.: University of Oxford (GBR), Borgonovi Francesca: OECD (FRA), Baert Stijn: Ghent University (BEL), Selwyn Neil: Monash University (AUS), Bers Marina Umaschi: Tufts University (USA), Ainscow Mel: University of Manchester (GBR), Soto Christopher J.: Colby College (USA) } \\
        \hline
\textbf{Social Science} & \textbf{Social Welfare} & 12484 & women, food, children, health, child, household, violence, households, gender, social, data, water, rural, farmers, countries & usa, gbr, chn, deu, aus, can, ind, nld, zaf, ita & World Bank, International Food Policy Research Institute, London School of Hygiene and Tropical Medicine, Johns Hopkins University, University of Oxford, UNICEF, University of North Carolina, Wageningen University, Michigan State University, Emory University \\
        \multicolumn{6}{|p{17.7cm}|}{\textbf{Top Authors:} McKenzie David: World Bank (USA), Headey Derek: International Food Policy Research Institute (USA), Qaim Matin: University of Göttingen (DEU), Barrett Christopher B.: Cornell University (USA), Ravallion Martin: Georgetown University (USA), Jayne Thomas S.: Michigan State University (USA), Klasen Stephan: University of Göttingen (DEU), Chandra-Mouli Venkatraman: World Health Organization (CHE), Krafft Caroline: St. Catherine University (USA), Hoddinott John: Cornell University (USA) } \\
        \hline

\end{tabular}
\label{tab:environmental_summary}
\end{table*}

\begin{table*}[ht]
\footnotesize
\centering
\begin{tabular}{|p{1.8cm}|p{2.5cm}|p{1.1cm}|p{3.2cm}|p{2.1cm}|p{4.7cm}|}
\hline
\textbf{Category} & \textbf{Cluster Name} & \textbf{\texttt{\#}Papers} & \textbf{Top TF-IDF Words} & \textbf{Top Countries} & \textbf{Top Institutions} \\
\hline
\hline

\textbf{Social Science} & \textbf{Urban Planning} & 10526 & climate, urban, adaptation, change, resilience, water, heat, cities, risk, green, local, social, migration, vulnerability, research & usa, gbr, aus, deu, chn, nld, can, swe, ita, zaf & Wageningen University, University of Melbourne, University College London, University of Oxford, Griffith University, World Bank, University of Queensland, University of Exeter, Arizona State University, University of Leeds \\
        \multicolumn{6}{|p{17.7cm}|}{\textbf{Top Authors:} Ford James D.: McGill University (CAN), Strobl Eric: University of Bern (CHE), McNamara Karen E.: University of Queensland (AUS), Noy Ilan: Victoria University of Wellington (NZL), Ide Tobias: Georg Eckert Institute for International Textbook Research (DEU), Biesbroek Robbert: Wageningen University (NLD), Wamsler Christine: Lund University (SWE), Sharifi Ayyoob: Hiroshima University (JPN), Robinson Stacy-ann: Colby College (USA), Kelman Ilan: University College London (GBR) } \\
        \hline
\textbf{Social Science} & \textbf{Data Science and AI} & 4620 & data, ai, public, digital, research, article, social, new, learning, big, work, smart, health, information, platform & usa, gbr, nld, aus, deu, chn, can, esp, ita, dnk & University of Oxford, University of Amsterdam, Stanford University, Utrecht University, Delft University of Technology, University of Copenhagen, University College London, London School of Economics, University of Toronto, University of Cambridge \\
        \multicolumn{6}{|p{17.7cm}|}{\textbf{Top Authors:} Floridi Luciano: University of Oxford (GBR), Taylor Linnet: Tilburg University (NLD), Mergel Ines: University of Konstanz (DEU), Kitchin Rob: Maynooth University (IRL), Ibsen Christian Lyhne: University of Copenhagen (DNK), Taddeo Mariarosaria: The Alan Turing Institute (GBR), Graham Mark: University of Oxford (GBR), Moynihan Donald P.: University of Wisconsin (USA), Kroll Alexander: Florida International University (USA), Lehdonvirta Vili: University of Oxford (GBR) } \\
        \hline
\textbf{Social Science} & \textbf{Vaccination} & 4158 & vaccine, vaccination, media, news, influenza, vaccines, health, covid, immunization, 19, hpv, social, coverage, political, information & usa, gbr, aus, che, can, chn, deu, nld, swe, ita & World Health Organization, Centers for Disease Control and Prevention, London School of Hygiene and Tropical Medicine, Johns Hopkins University, University of Oxford, University of Gothenburg, National Center for Immunization and Respiratory Diseases, University of Amsterdam, University of Toronto, University of Washington \\
        \multicolumn{6}{|p{17.7cm}|}{\textbf{Top Authors:} van der Linden Sander: University of Cambridge (GBR), Thompson Kimberly M.: Kid Risk (USA), Rand David G.: Massachusetts Institute of Technology (USA), Pennycook Gordon: University of Regina (CAN), Tandoc Edson C.: Nanyang Technological University (SGP), Shehata Adam: University of Gothenburg (SWE), Nyhan Brendan: Dartmouth College (USA), Lewandowsky Stephan: University of Bristol (GBR), Garnett Holly Ann: Royal Military College of Canada (CAN), James Toby S.: University of East Anglia (GBR) } \\
        \hline

\end{tabular}
\label{tab:environmental_summary}
\end{table*}

\clearpage

\begin{figure}[!h]
\centering
\includegraphics[width=0.8\textwidth]{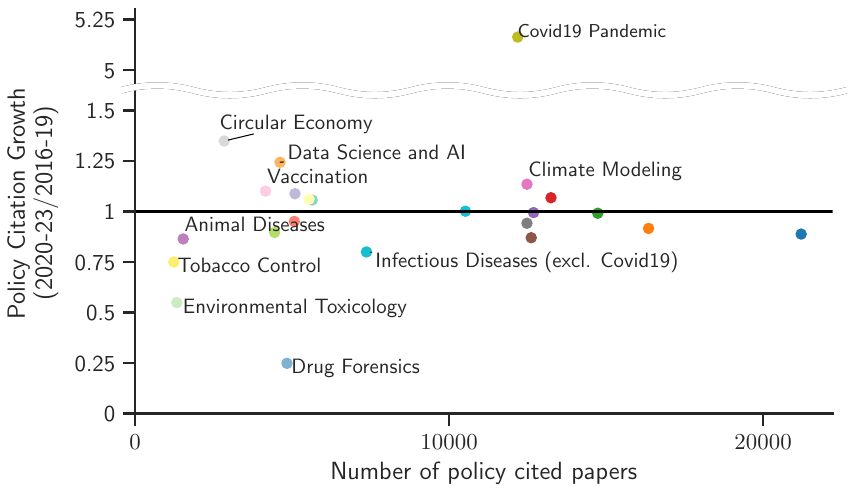}
  \caption*{\textbf{Figure S2. Policy citation growth rates versus total policy-cited papers by domain} Growth trajectories of policy citations across scientific domains, plotting growth rates (the share of policy-cited papers in 2020-2023 relative to the share in 2016-2019) against absolute citation counts. Growth rate = 1 indicates no change; values above/below indicate increasing/decreasing policy attention. The Covid-19 Pandemic cluster exhibits both high volume and rapid growth, reflecting urgent policy attention during the global health crisis. Emerging fields like Circular Economy and Data Science and AI show positive growth despite smaller absolute numbers. Conversely, Environmental Toxicology and Drug Forensics display below-average growth, with Drug Forensics falling below unity, indicating declining policy relevance. These divergent trajectories highlight how policy attention shifts across scientific domains in response to global challenges.}
\end{figure}
\clearpage


\begin{figure}[htbp]
  \centering
  \includegraphics[width=\textwidth]{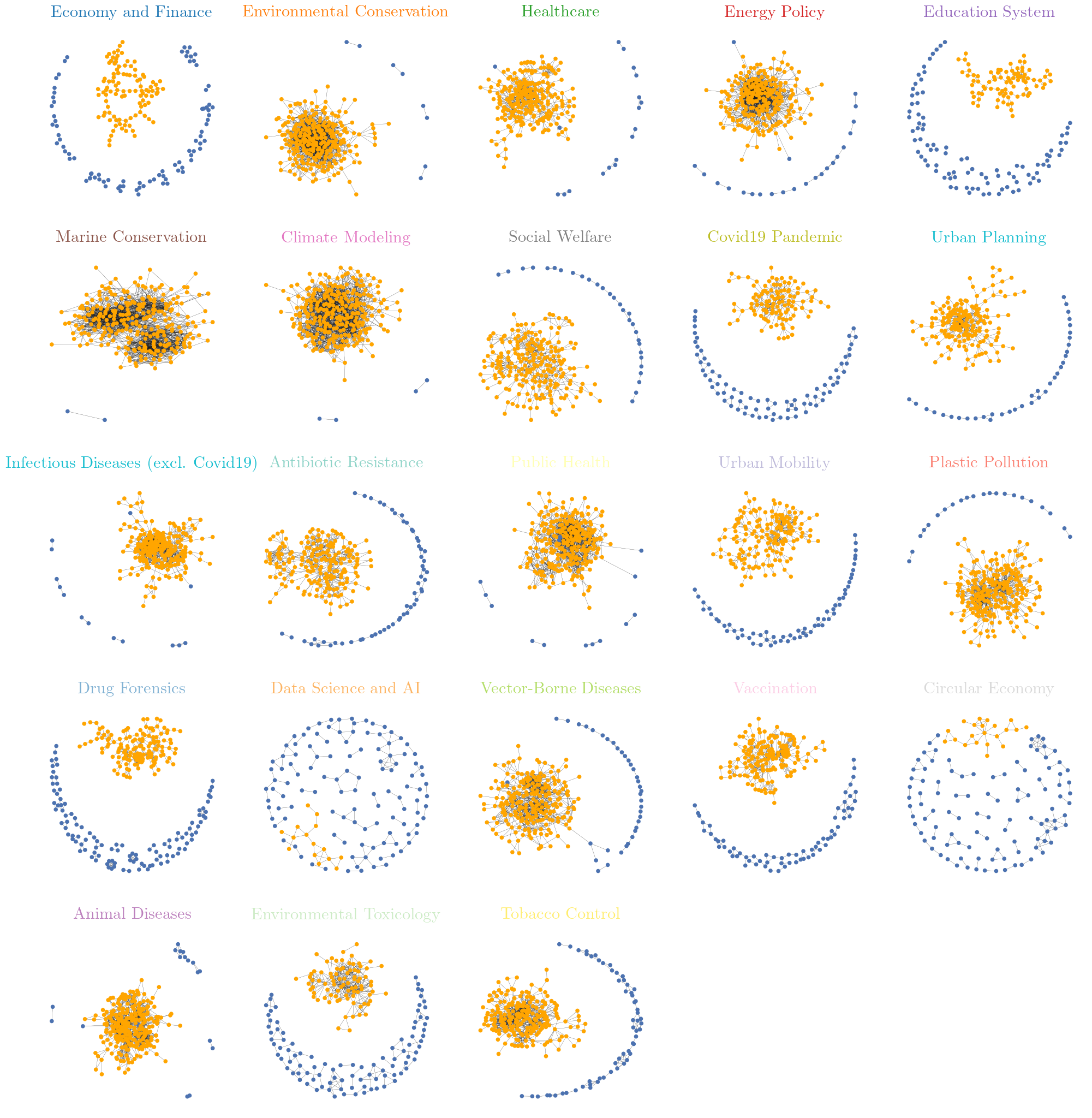}
  \caption*{\textbf{Figure S3. Co-authorship networks of top 300 policy-cited scientists across 23 domains} Orange nodes represent scientists in the largest connected component, indicating core collaborative groups; blue nodes show peripheral scientists. Each edge represents at least one co-authored paper. Network topology varies dramatically across fields. Climate Modeling and Public Health display dense, centralized structures with dominant cores, reflecting established research hierarchies and coordinated policy influence. In contrast, Data Science and AI and Education System exhibit fragmented networks with multiple smaller components, suggesting distributed expertise and emerging policy connections. These structural differences reveal how disciplinary organization shapes pathways from science to policy.}
  \label{fig:s3}
\end{figure}

\clearpage



\begin{table}[h!]
\centering
\caption*{\textbf{Table S4. List of IPCC Reports used to identify overlap between policy-influential scientists and formal policy advisors.} To assess overlap between policy-influential scientists and formal policy advisors, we compiled a comprehensive roster of IPCC authors from 12 assessments and special reports (2011-2023). Author names were extracted from PDFs of the reports, standardized to a "surname, given name" format, and de-duplicated, yielding 1,639 unique IPCC authors. This roster was matched against Climate Modeling PI-Sci using fuzzy string matching (case-insensitive). 97 PI-Sci were identified as authors on at least one of the selected IPCC reports. The table lists report titles, publication years, and number of authors for each IPCC assessment used in this analysis.}
\begin{tabular}{m{10cm}>{\centering\arraybackslash}m{2.5cm}>{\centering\arraybackslash}m{2.5cm}}
\toprule
\centering \textbf{Report} & \textbf{Publication Year} & \textbf{Number of Authors} \\ \midrule
AR6 Synthesis Report & 2023 & 30 \\
AR6 - The Physical Science Basis (AR6-WG1) & 2023 & 234 \\
AR6 - Impacts, Adaptation, and Vulnerability (AR6-WG2) & 2023 & 270 \\
AR6 - Mitigation of Climate Change (AR6-WG3) & 2023 & 239 \\
Special Report on the Ocean and Cryosphere in a Changing Climate (SROCC)& 2019 & 103 \\
Special Report on Global warming of 1.5°C (SR15) & 2018 & 91 \\
Special Report on Climate Change and Land (SRCCL) & 2019 & 107 \\
AR5 - The Physical Science Basis (AR5-WG1) & 2014 & 259 \\
AR5 - Impacts, Adaptation, and Vulnerability (AR5-WG2) & 2014 & 303 \\
AR5 - Mitigation of Climate Change (AR5-WG3) & 2014 & 287 \\
Special Report on Managing the Risks of Extreme Events and Disasters to Advance Climate Change Adaptation (SREX) & 2012 & 103 \\
Special Report on renewable energy sources and climate change mitigation (SRREN) & 2011 & 158 \\ \hline
\textbf{Total Number of Unique Authors} &  & 1639 \\
\bottomrule
\end{tabular}
\label{table:ipcc_reports}
\end{table}



\end{document}